\documentclass[conference]{IEEEtran}

\usepackage{helvet, courier, graphicx}
\usepackage{subfig}
\usepackage{wrapfig}
\usepackage[colorlinks,urlcolor=blue,citecolor=blue,linkcolor=blue]{hyperref}
\usepackage{color}
\usepackage{placeins}
\usepackage{amsmath}
\IEEEoverridecommandlockouts
\DeclareCaptionLabelSeparator{nobreakspace}{\nobreakspace}

\usepackage{tabularx}
\newcommand{\squeezeup}{\vspace{-.5mm}}
\begin{document}
%
\title{Understanding Co-evolution in Large Multi-relational Social Networks }

\author{\IEEEauthorblockN{Ayush Singhal}
\IEEEauthorblockA{Computer Science and\\Engineering\\
University of Minnesota\\
Minnesota, USA\\
Email: singh196@umn.edu}
\and
\IEEEauthorblockN{Atanu Roy}
\IEEEauthorblockA{Computer Science and\\Engineering\\
University of Minnesota\\
Minnesota, USA\\
Email: royxx215@umn.edu}
\and
\IEEEauthorblockN{Jaideep Srivastava}
\IEEEauthorblockA{Computer Science and\\Engineering\\
University of Minnesota\\
Minnesota, USA\\
Email: srivasta@umn.edu}
}

%
\maketitle

\begin{abstract}
Understanding dynamics of evolution in large social networks is an important problem.  In this paper, we characterize evolution in large multi-relational social networks. The proliferation of online media such as Twitter, Facebook, Orkut and MMORPGs\footnote{Massively Multi-player Online Role Playing Games} have created social networking data at an unprecedented scale. Sony's Everquest 2 is one such example. We used game multi-relational networks to reveal the dynamics of evolution in a multi-relational setting by macroscopic study of the game network. Macroscopic analysis involves fragmenting the network into smaller portions for studying the dynamics within these sub-networks, referred to as `communities'. From an evolutionary perspective of multi-relational network analysis, we have made the following contributions. Specifically, we formulated and analyzed various metrics to capture evolutionary properties of networks. We find that co-evolution rates in trust based `communities' are approximately $60\%$ higher than the trade based `communities'. We also find that the trust and trade connections within the `communities' reduce as their size increases. Finally, we study the interrelation between the dynamics of trade and trust within `communities' and find interesting results about the precursor relationship between the trade and the trust dynamics within the `communities'. 


\end{abstract}


%

\section{Introduction}

Understanding evolution of communities \cite{xie2011overlapping, palla2007quantifying, ramchand2012co} have gained importance over the past decade. Evolution of communities help researchers identify the evolving nature of human socialization. The huge amount of social network data at our perusal has fueled the research in this direction. Primarily research is performed on uni-relational networks, which refers to networks having a single set of edges. Most of the networks dataset that we get hold of \footnote{https://snap.stanford.edu/data/} represent this kind of networks. But we are starting to have diverse kinds of datasets. One of these diverse datasets includes multi-relational data, where instead of a single set of edges, we have multiple set of edges. The introduction of multiple sets of edges introduces an intriguing problem in the context of evolution of communities: co-evolution. Earlier the researchers did not have to deal with multiple relations and thus it was enough to handle evolution of communities based on a single relation. But with the advent of datasets containing multiple relations, arose the opportunity to study co-evolution of communities in multiple relations. Multi-relational datasets are very hard to get a hold of, but the way social network data is increasing, we can assume that these relations will be available in abundance in the near future.

As discussed earlier, the availability of multi-relational datasets is not very high. Very few researchers have based their investigations on these kinds of datasets~\cite{berlingerio2011finding, carchiolo2011communities, cai2005community}. One of them is Cai \textit{et. al.} in \cite{cai2005community} where they have looked at the problem of community mining in multi-relational datasets. On the other hand, the present work addresses the problem of measuring co-evolution dynamics in communities. To start off, the networks are partitioned into communities, using well known community detection algorithms \cite{palla2007quantifying} and important statics are derived from them. Next we look at the dynamics inside these communities discovered by the algorithms. Once we have understood the evolution in communities, we move our investigation to quantitatively study co-evolution of communities. Finally, the direction of influence in the multi-relational network is determined by statistical exploration of the interplay between the ``interesting'' events within the communities.

In this work, we find that the smaller sized communities have higher connectivity in terms of the trade and the trust links. The experiments on co-evolution show that the trust based communities exhibit $60\%$ higher co-evolution rate than trade based communities. We also find that within the tightly knit communities, the occurrence of unusual trade dynamics is followed by unusual trust dynamics. This observation reveals several insights about the various dynamics occurring within a large scale multi-relational network. 

The main contributions of this work are summarized in the following manner:

\begin{itemize}
\item We formulated and compared metrics to study evolution in multi-relational network. 

\item We find that evolution and co-evolution metrics are strongly dependent on the type of community. Specifically, we find that communities based on the trust relationship have higher evolution and co-evolution rates than the trade based communities.
\item We find that the dynamics within a community are influenced by its size. Specifically, we find that the rate of increase in trust in communities decreases as the community size increases.
\item Finally, we find precursors of various dynamics within the communities. For example, unusual events in the trust dynamics are preceded by unusual trade dynamics. 

\end{itemize}

\section{Related Works}

Understanding large networks and community detection in those networks is a very well researched topic \cite{leskovec2009meme, leskovec2010empirical,cai2005community}. Most of the research that have been done in this field is performed on uni-relational networks \cite{xie2011overlapping, leskovec2010empirical, palla2007quantifying}. In one of the works by Borabora et. al \cite{DBLP:conf/saso/BorboraAHSW11}, multi-relational networks are studied in the area of computational trust. The primary objective of the research was to identify robust predictors of trust in an online virtual environment. In a different work \cite{cai2005community}, Cai et al investigate community detection (community mining) but the study primarily focused on how to perform efficient detection of communities in a multi-relational setting. Various aspects of communities have already been studied extensively in various domains including evolution of communities \cite{bradshaw1986evolution, greene2010tracking, palla2007quantifying}. As discussed earlier most of these analyzes were performed in uni-relational networks. Thus the problem of co-evolution of the multiple relations in a network within the communities is not addressed in these literature. 

Researchers in the past decade have extensively investigated the problem of finding influencers in a social network. The seminal work by Kempe \textit{et. al.} \cite{kempe2003maximizing} in 2003 defined the problem of finding influencers who can maximize the flow of information in a network. Various researchers have investigated various methods to maximize influence in a network \cite{chen2009efficient} for finding correlation and influence in social networks \cite{anagnostopoulos2008influence}. Kempe \textit{et. al.} \cite{kempe2003maximizing} in their seminal work assumed that the influence capability between nodes are already provided to the problem. Subbian \textit{et. al.} in \cite{subbian2013} have moved away from the approach of assuming influence weights and have instead calculated social capital inside a network to find influencers. 

\section{Data used}
Sony EverQuest II(EQ II) game provides an online environment where multiple players can log in and coordinate with each other to achieve a particular mission. The game provides several mechanisms such as chat and e-mail for instantaneously interaction. We used the server logs from this game, and we extracted the information needed for our experiments from these logs in which the players perform various interactions with each other. In this section, we describe the networks used in our experiments. 

\subsection{Trust Network}
In EQ II the players are limited by the number of items they can carry at a time, players buy houses as a temporary storage to retain their weapons and other accessories. Players have the ability to share their house access with other players. The network thus formed in the process is referred to as the trust network. We have 9 months of data from Jan-01-2006 to Sep-11-2006 with 51,428 nodes and 72,446 edges where nodes represent player characters in the game and edges represent a in game character giving another in game character permission to access his or her house. Each edge has a time stamp when the access was granted. 
%
%
 
\subsection{Trade Network}
Like the real world, in EQ II players can exchange goods for coins or other goods. The exchange of items between two in game characters lead to a formation of trade links. We analyzed such a network which contains 181,488 nodes and 22,24,450 edges over a period of 9 months from Jan-01-2006 and Sep-11-2006.

\begin{table}
	\centering
		\caption{Table showing statistics about the various overlaps between networks.}
		\scalebox{0.75}{
		\begin{tabular}{|l|l|l|l|}
			\hline
			 &Trade & Reduced trade & Total nodes\\
			\hline
			Trust & 44226 (86\%)  & 27005 (52.5\%) & 51428\\
			&(24.4\%) &(46.6\%)&\\
			\hline
			 Total nodes & 181488 & 57935&\\
			\hline
		\end{tabular}
	}
	\label{tab:showingStatisticsAboutNetwork}
	\squeezeup
\end{table}

\section{Definitions}

\subsection{Clique Percolation Method (CPM)}

Clique percolation method \cite{palla2007quantifying} or CPM is a community detection algorithm to detects communities from $k$ cliques. CPM uses adjacent cliques to build up a community. Two cliques are considered adjacent according to CPM if they share $k-1$ edges. A community is defined as a maximal union of adjacent cliques. $k$ is an user defined parameter and for this paper we empirically found that CPM performs best when $k=3$.

\subsection{Clauset Newman and Moore Algorithm (CNM)}
Clasuet Newman and Moore algorithm \cite{clauset2004finding} or CNM proposed by Clauset \textit{et. al.} is a community detection algorithm based on the modularity property \cite{newman2004finding} of a network. It uses a variation of the hierarchical based clustering in which the principal objective is to maximize the edges inside a community and minimize inter-community edges in a network. 

\section {Experiments}
The following are the description of the various notations used in computing metrics for different experiments.

Let $V = \{v_1, \ldots, v_n\}$ be the set of all nodes.
 
Let $E = E_{trust} \bigcup E_{trade} $ be the set of all links. $E_{trust}$ represents all the links belonging to the trust relation and $E_{trade}$ refers to the links belonging to the trade relation. The set $E$ is a multi-set, i.e., edges between the same set of actors can occur more than once. Moreover $E^t$ refers to all links that exist at time $t$.
For each link $e \in E$, we assume that we can perform the following actions: 
 

$src(e)$ = source node of link $e$

$dst(e)$ = destination node of link $e$

Moreover we can perform the following actions on the multi-relational network: 
 
$E_{\rm trust}^t =  E_{\rm trust}  \cap E^t$

$E_{\rm trade}^t = E{\rm trade} \cap E^t$

${C_k}$ represents communities where $k=3$

Since CNM does not allow overlapping communities, we have $C_{k_1} \cap C_{k_2} = \emptyset$ ; $ \forall k_1 \neq k_2$
 
An edge within the community $k$ at time $t$ is defined as follows:
 
$E_{\rm trade}^{t,k} = \{ e \in E_{\rm trade}^t | src(e) \in C_k$ and $dst(e) \in C_k \}$

$E_{\rm trust}^{t,k} = \{ e \in E_{\rm trust}^t | src(e) \in C_k$ and $dst(e) \in C_k \}$

An edge peripheral to the community $k$ at time $t$ is defined as follows:
 
$P_{\rm trade}^{t,k} = \{ e \in E_{\rm trade}^t | (src(e) \in C_k$ and $dst(e) \notin C_k$) or ($dst(e) \in C_k$ and $src(e) \notin C_k)\}$

$P_{\rm trust}^{t,k} = \{ e \in E_{\rm trust}^t | (src(e) \in C_k$ and $dst(e) \notin C_k$) or ($dst(e) \in C_k$ and $src(e) \notin C_k)\}$

\begin{table}
	\centering
	\caption{Table showing the counts of communities derived from $G_{trust }$ and $G_{trade}'$ usign CNM and CPM algorithms.}
	\scalebox{0.75}{
		\begin{tabular}{|l||l|l|}
		\hline
		Community type & CNM(size$>3$) based count & CPM(k=3) based count\\
		\hline
		\hline
		Trust-communities & 2889 & 2876\\
		\hline
		Trade-communities & 2073 & 2898\\
				\hline
			
		\end{tabular}
		}
	\label{tab:community_Count}
	\squeezeup
\end{table}

\subsection{Fragmenting network into communities}
Given the multi-relational network $G=<V,E>$, we have used $G_{trust}=<V,E_{trust}>$ and $G_{trade}=<V,E_{trade}>$ networks to construct two types of communities. Using the $G_{trust}$, we derive communities based on trust relationship between the nodes and using the $G_{trade}$, we derive communities based on the trade relationship between the nodes. In order to detect communities in $G_{trust}$ network, the entire network was viewed as a snapshot for the entire observation period. The direction of the edges was removed and the weights on the edges were dropped. Then the communities were detected on this unweighted, undirected network using two community definitions: CNM and CPM. 

In the trade ($G_{trade}$) network, the communities were identified on a reduced network $G_{trade}'$ ( as shown in Table~\ref{tab:showingStatisticsAboutNetwork}). Note that the reduction in network is not performed for the trust network and thus $G_{trust}'$ does not exist. Since the trade links, unlike trust links, signify instant interactions rather than a long term relationship, we derive a new trade network $G_{trade}'$ where the links can be considered better proxy. However, this new network is used only for the purpose of fragmenting the $G_{trade}$ network in meaningful communities. While accounting for the various kinds of trade dynamics and trade-trust interplay, we have used the original trade network, $G_{trade}$. In order to derive the $G_{trade}'$ from $G_{trade}$, firstly, all the edges in $G_{trade}$ are made undirected. By observing the $G_{\rm trade}$ as a snapshot for the entire observation period, edges in $G_{trade}$ are assigned weights equal to the frequency of appearance of an edge in the entire observation period. The weights($w$) on these edges correspond to the strength of interaction. We established a threshold of $5$ interactions between actors to be represented in the new graph which will be represented as  $G_{\rm trade}'$ throughout the rest of the paper. The edges having a weight of $< 5$ are ignored. Finally, the weights on the edges in $G_{trade}'$ are dropped. Having derived a unweighted, undirected network $G_{\rm trade}'$, the communities are defined on this network using CNM and CPM community detection algorithms.

 
\subsection{Statistics for Multi-relational Network}
In this section, we discuss some of the statistical properties of the communities derived in the previous experiment. For the sake of convenience, we refer to the communities derived from the $G_{trust}$ as trust communities and those derived from $G_{trade}'$ as trade communities. However, the communities are only a set of nodes which can have both trust and trade edges. 

\begin{figure}
	\centering
		\subfloat[]{\includegraphics[width=0.25\textwidth]{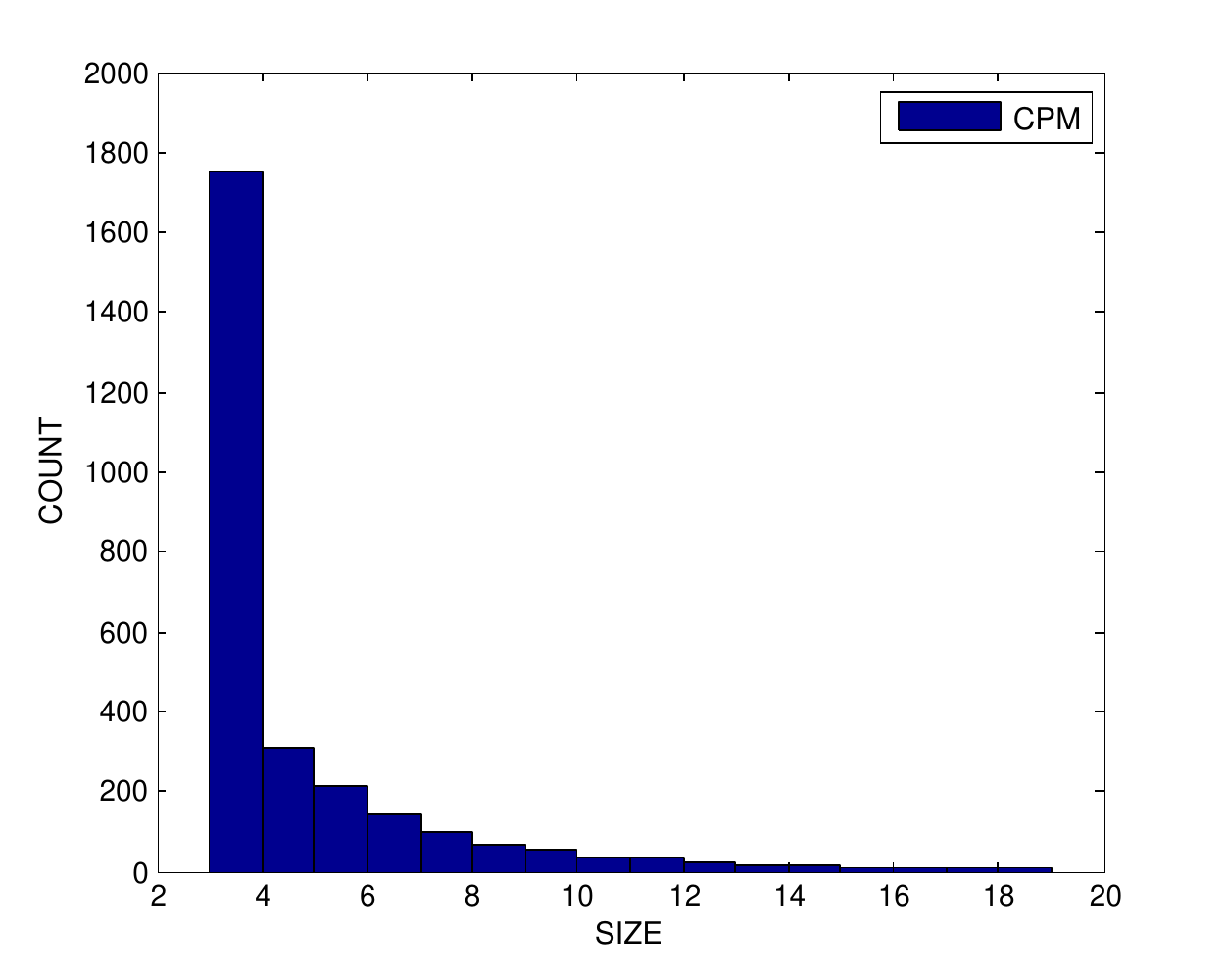}}~\subfloat[]{\includegraphics[width=0.25\textwidth]{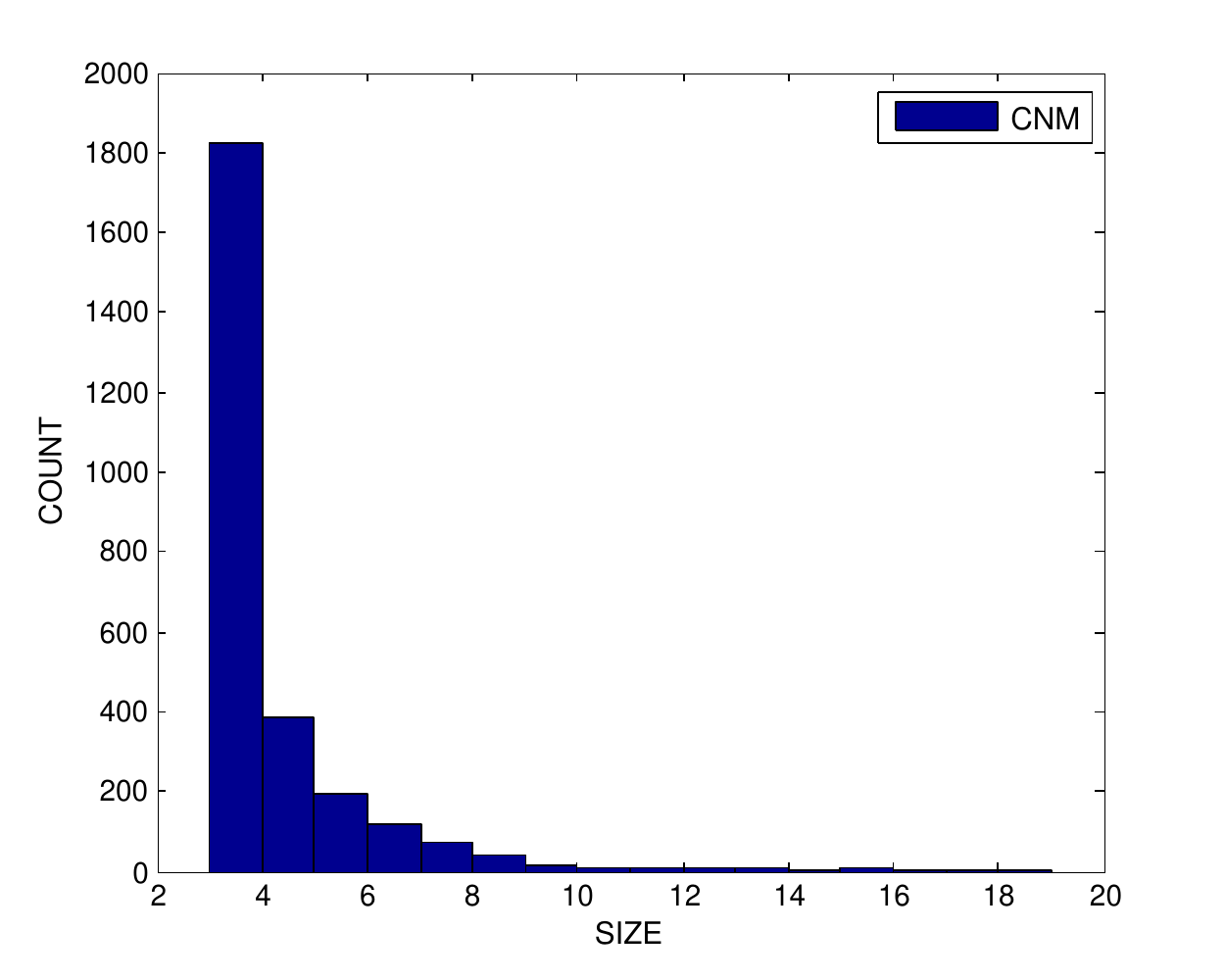}}
	\caption{Plots showing size distribution of the communities generated by CPM and CNM respectively.}
	\label{fig:trust_inter_intra}
	\squeezeup
\end{figure}

\begin{figure}
	\centering
		\subfloat[]{\includegraphics[width=0.25\textwidth]{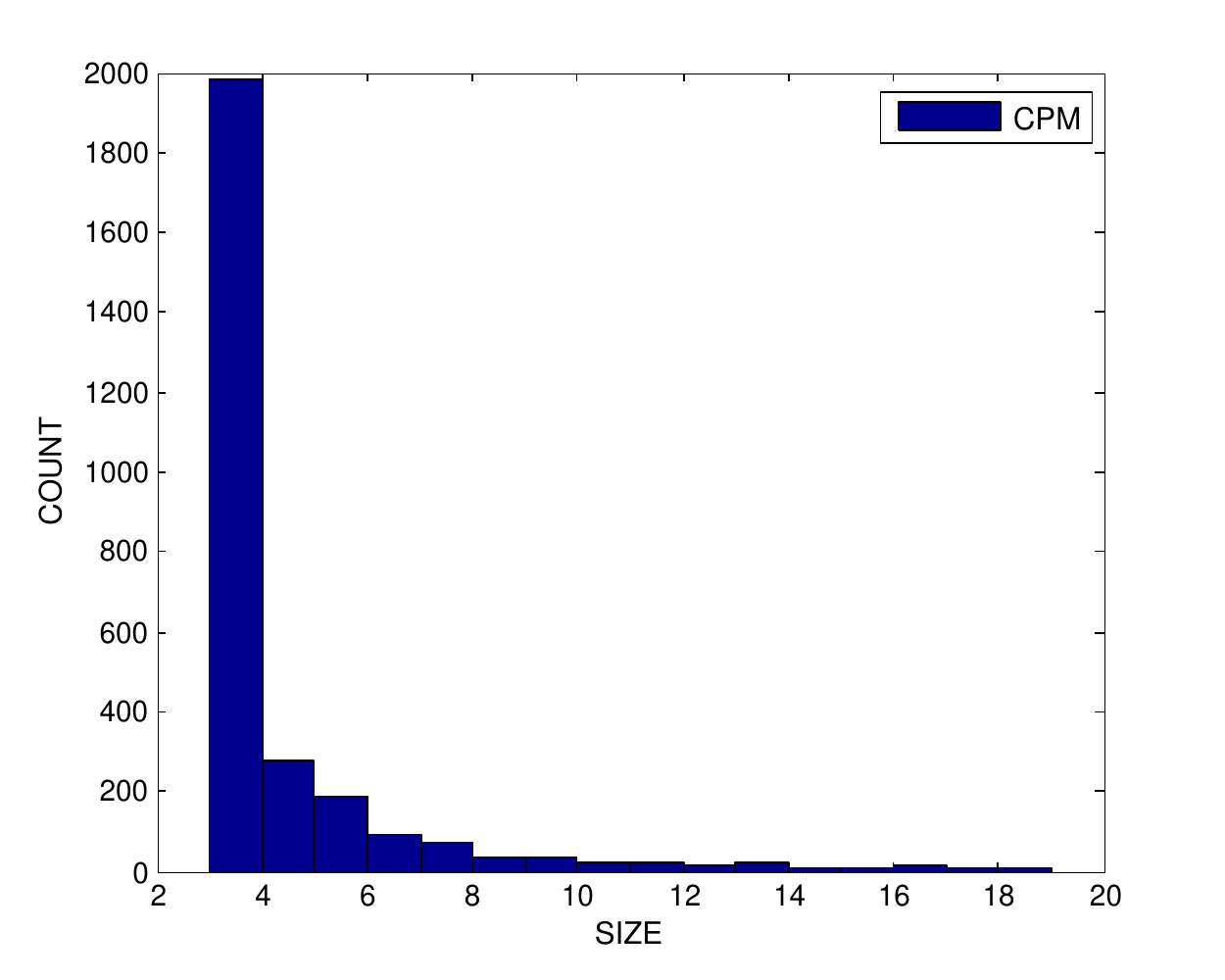}}~\subfloat[]{\includegraphics[width=0.25\textwidth]{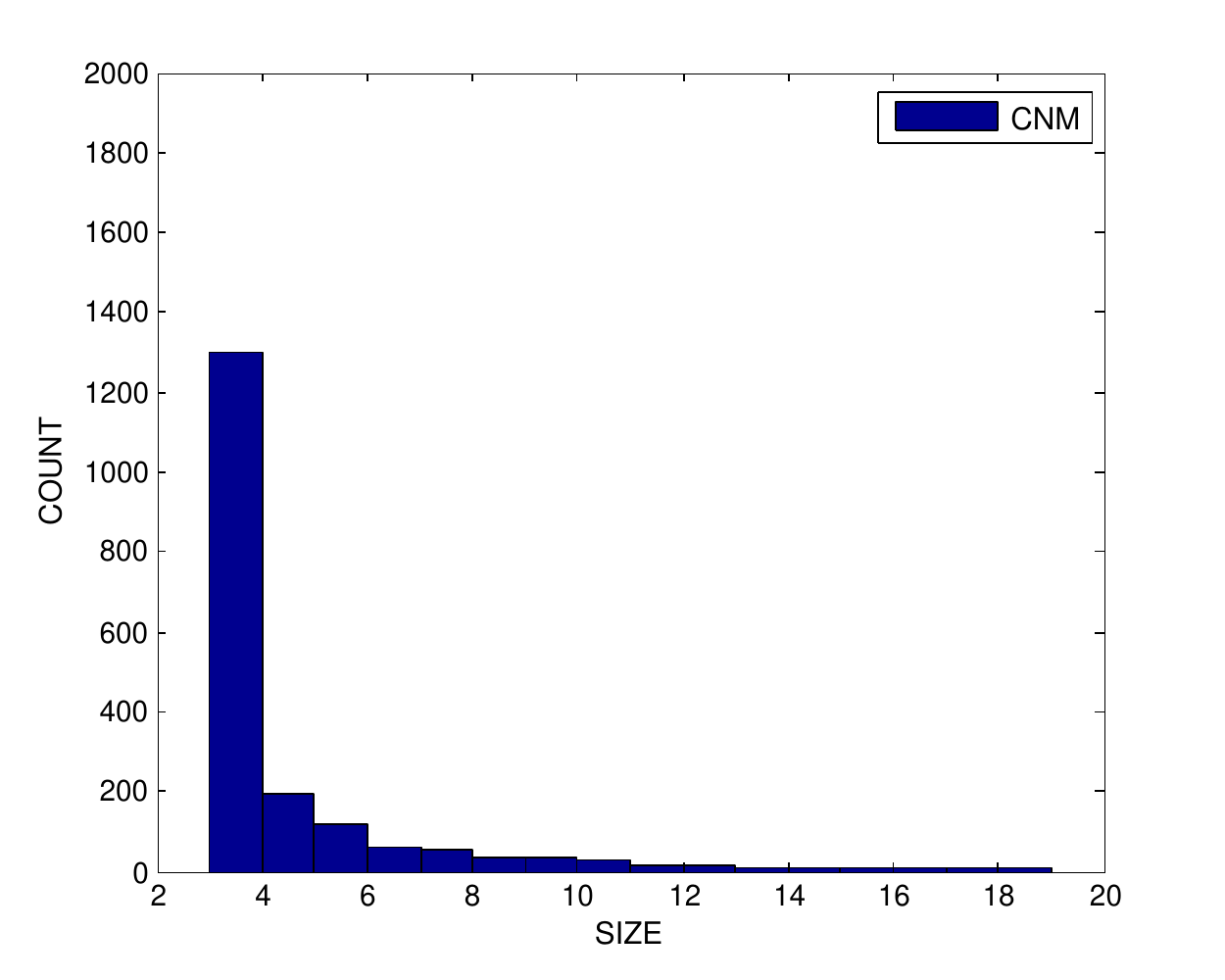}}
	\caption{Plots showing size distribution of the communities generated from the trade network using CPM and CNM.}
	\label{fig:trade_inter_intra}
	\squeezeup
\end{figure}
\begin{table}
	\centering
	\caption{Table showing the size of various grouping of trust communities based on sizes.   }
	\scalebox{0.8}{
		\begin{tabular}{|l|l|l|l|}
		\hline
		Trust communities  & Group 1 &  Group 2 & Group 3  \\
		 & (size:[3]) & (size:[4,10)) & (size:[10-)) \\
		
		\hline
		\hline
		CNM based (2889) & 1154(39.9\%) &  1482(51.3\%) & 253(8.8\%) \\
		\hline
		CPM based (2876) & 1242(43.2\%)  & 1346(46.8\%) & 288(10\%) \\
		\hline

		\end{tabular}
		}
	
	\label{tab:Table grouping stats}
	\squeezeup
\end{table}

Table~\ref{tab:community_Count} gives the description of the number of communities obtained on $G_{trust}$ and $G_{trade}'$ using CPM and CNM definitions. As shown in table~\ref{tab:community_Count}, the number of communities obtained using CNM is comparable to the number of communities obtained using CPM algorithms. The difference is due to the incomplete coverage of the network using CPM definition for communities.
\begin{table}
	\centering
	\caption{Table showing the size of various grouping of trade communities based on sizes.   }
	\scalebox{0.75}{
		\begin{tabular}{|l|l|l|l|}
		\hline
		Trade communities  & Group 1 &  Group 2 & Group 3  \\
		 & (size:[3]) & (size:[4,10)) & (size:[10-)) \\
		
		\hline
		\hline
		CNM based (2073) & 899(43.3\%) &  869(42.0\%) & 305(14.7\%) \\
		\hline
		CPM based (2898) & 1412(48.7\%)  & 1246(43.0\%) & 240(8.3\%) \\
		\hline

		\end{tabular}
		}
\label{tab:Table trade grouping stats}
\squeezeup
\squeezeup
\end{table}

Figures~\ref{fig:trust_inter_intra} and ~\ref{fig:trade_inter_intra} show the distribution of community sizes for the trust and trade communities respectively. Figures~\ref{fig:trust_inter_intra}(a) and ~\ref{fig:trade_inter_intra}(a) show the distribution when the communities are defined using the CPM algorithm and figures~\ref{fig:trust_inter_intra}(b) ~\ref{fig:trade_inter_intra}(b), show the distribution when the communities are defined using the CNM. The large variance in sizes of communities motivates us to analyze the various dynamics of communities by grouping them based on their sizes. The grouping of communities into different categories helps in a generic characterization of the various metrics based on the size of the communities. Thus, using the information from the size distribution, we categorized the communities as shown in tables~\ref{tab:Table grouping stats} and ~\ref{tab:Table trade grouping stats} for trust and trade communities respectively. 

Tables~\ref{tab:Table grouping stats} and ~\ref{tab:Table trade grouping stats} provide the count of communities in each of the categories. In both the tables, the first category(group 1) contains communities which have size $= 3$. As shown in the table, group 1 communities form a large proportion of the total communities (approximately $40-48\%$) and are thus kept in a separate category. The second category(group $2$) contains communities which have sizes ranging from $4$ to $9$. As shown in the table, the group $2$ covers approximately $42-50\%$ of the total communities. Finally, the third category(group $3$) comprises of communities which have size greater than equal to $10$. This group covers only approximately $10\%$ of total communities.

		%
		%
			%
	%

\subsection{Analysis of community evolution}
In this section, we discuss the experimental design for the analysis of the community evolution in a multi-relation network. There are multiple aspects of evolution of a community. We consider the intra-community structure evolving over time. In order to track the intra-community structure over time, we define a metric called \textit{connectivity}. Connectivity can be defined as a measure of the intra-connectedness of a community. At any given snapshot of time($t$), the connectivity($Q$) of a community($k$) can be formally defined as follows:

$Q_{\rm trade}^{t,k}=\frac{|E_{\rm trade}^{t,k}|}{|C_k|\times(|C_k|-1)}$ and

$Q_{\rm trust}^{t,k}=\frac{|E_{\rm trust}^{t,k}|}{|C_k|\times(|C_k|-1)}$
\begin{figure}
	\centering
		\includegraphics[width=0.33\textwidth]{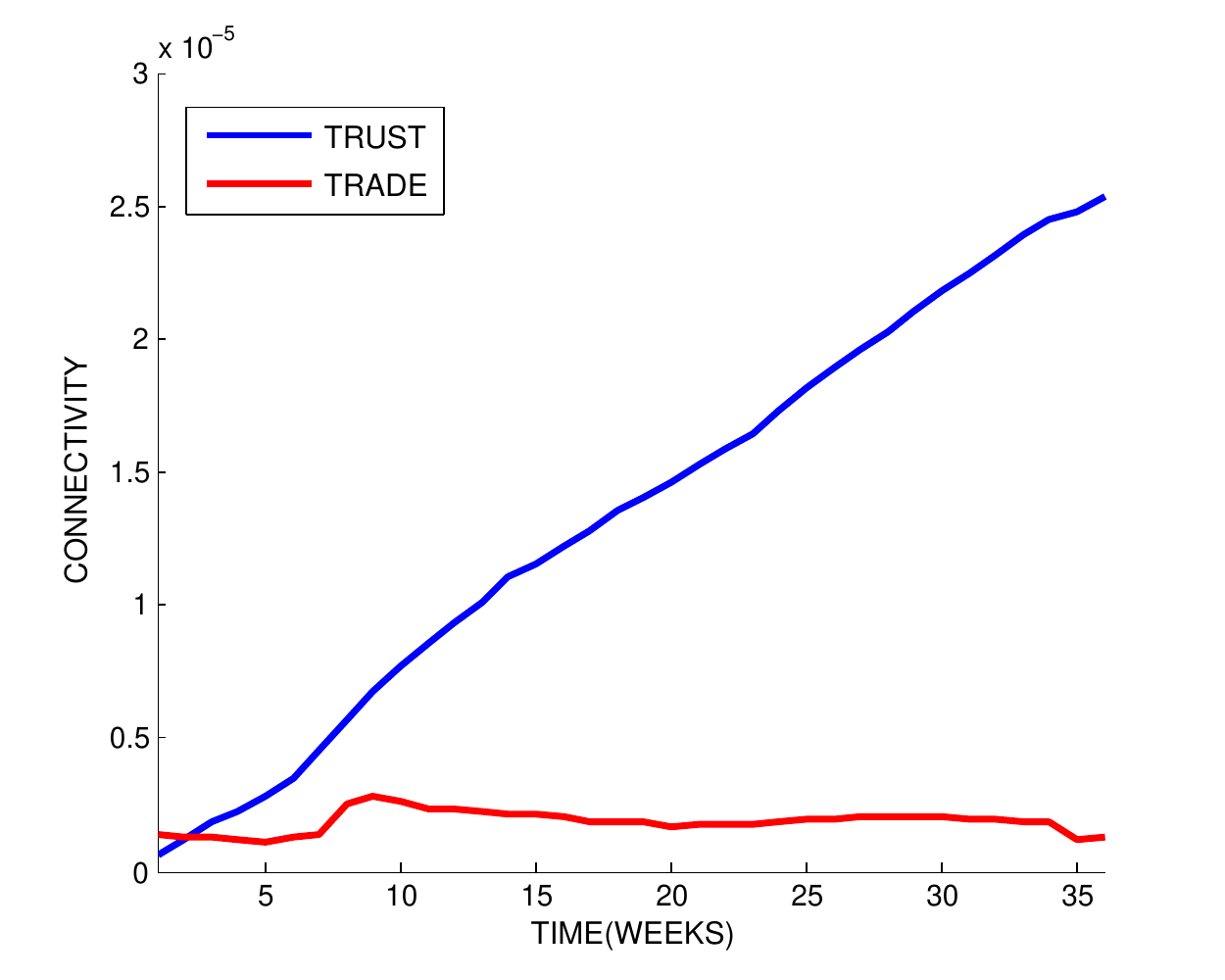}
	\caption{Evolution of overall trade and trust connectivity in the network.}
	\label{fig:overall_network_connectivity_trust_trade}
	\squeezeup
\end{figure}

\begin{figure*}[t]
	\centering
		\subfloat[]{\includegraphics[width=0.33\textwidth]{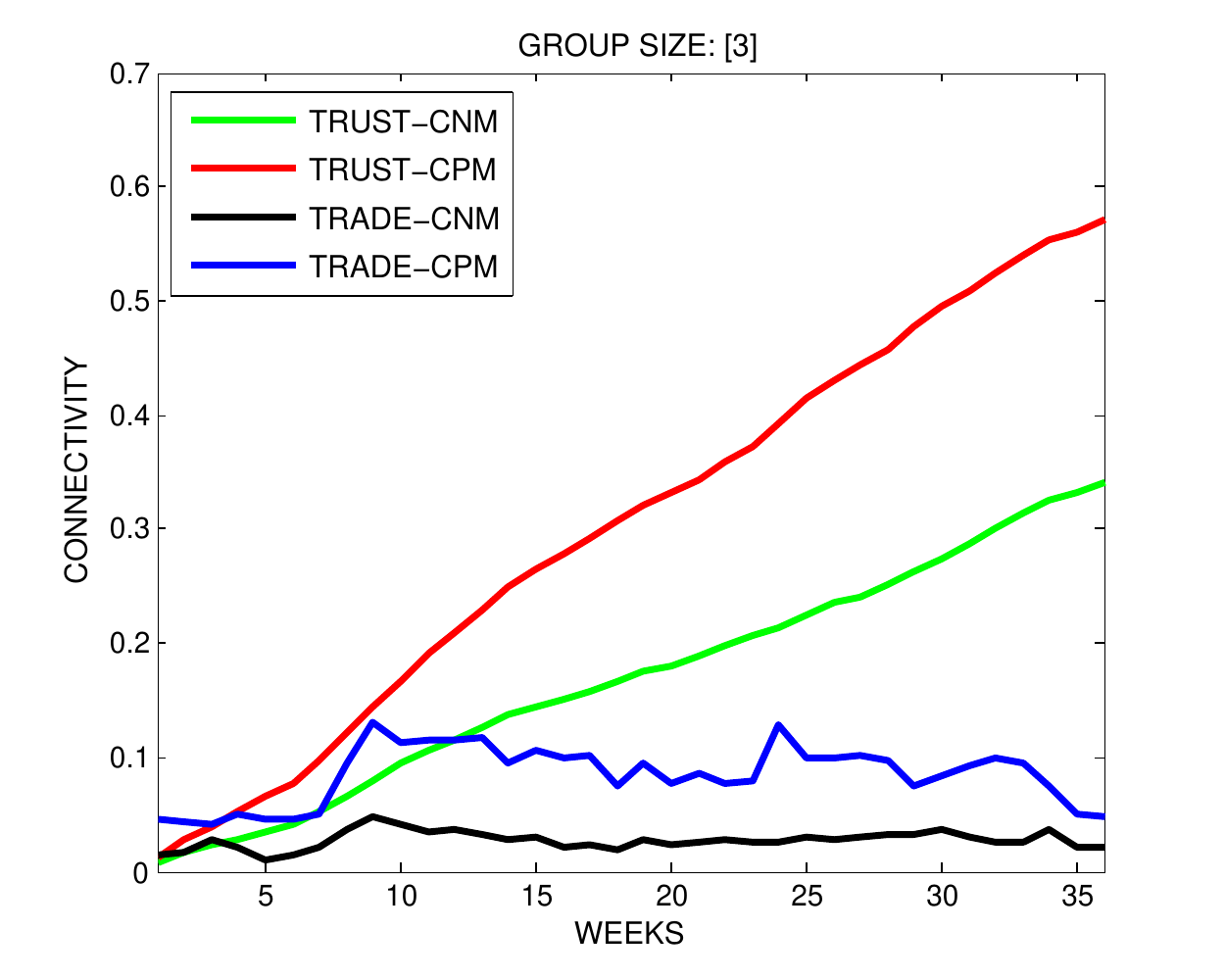}}~\subfloat[]{\includegraphics[width=0.33\textwidth]{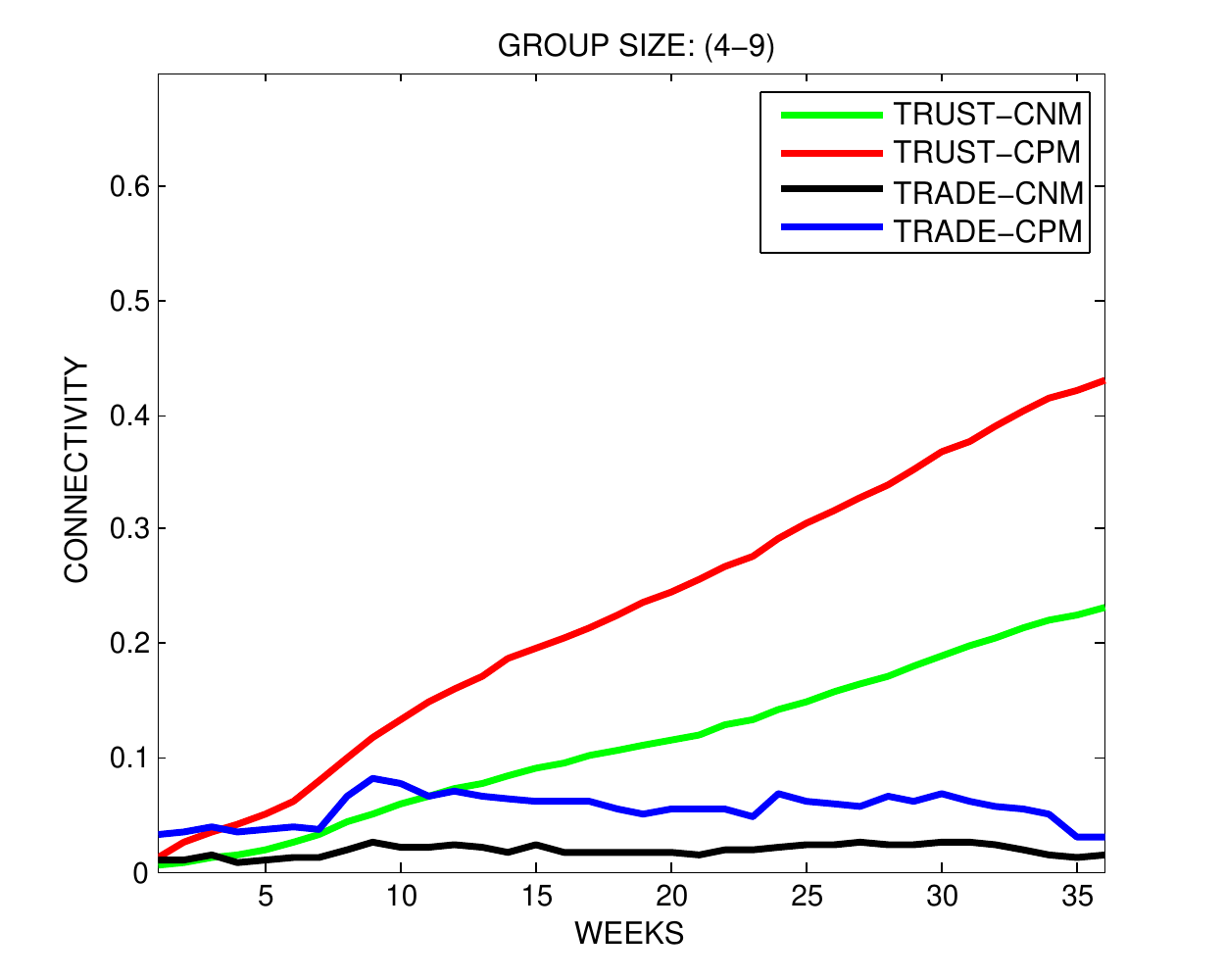}}~\subfloat[]{\includegraphics[width=0.33\textwidth]{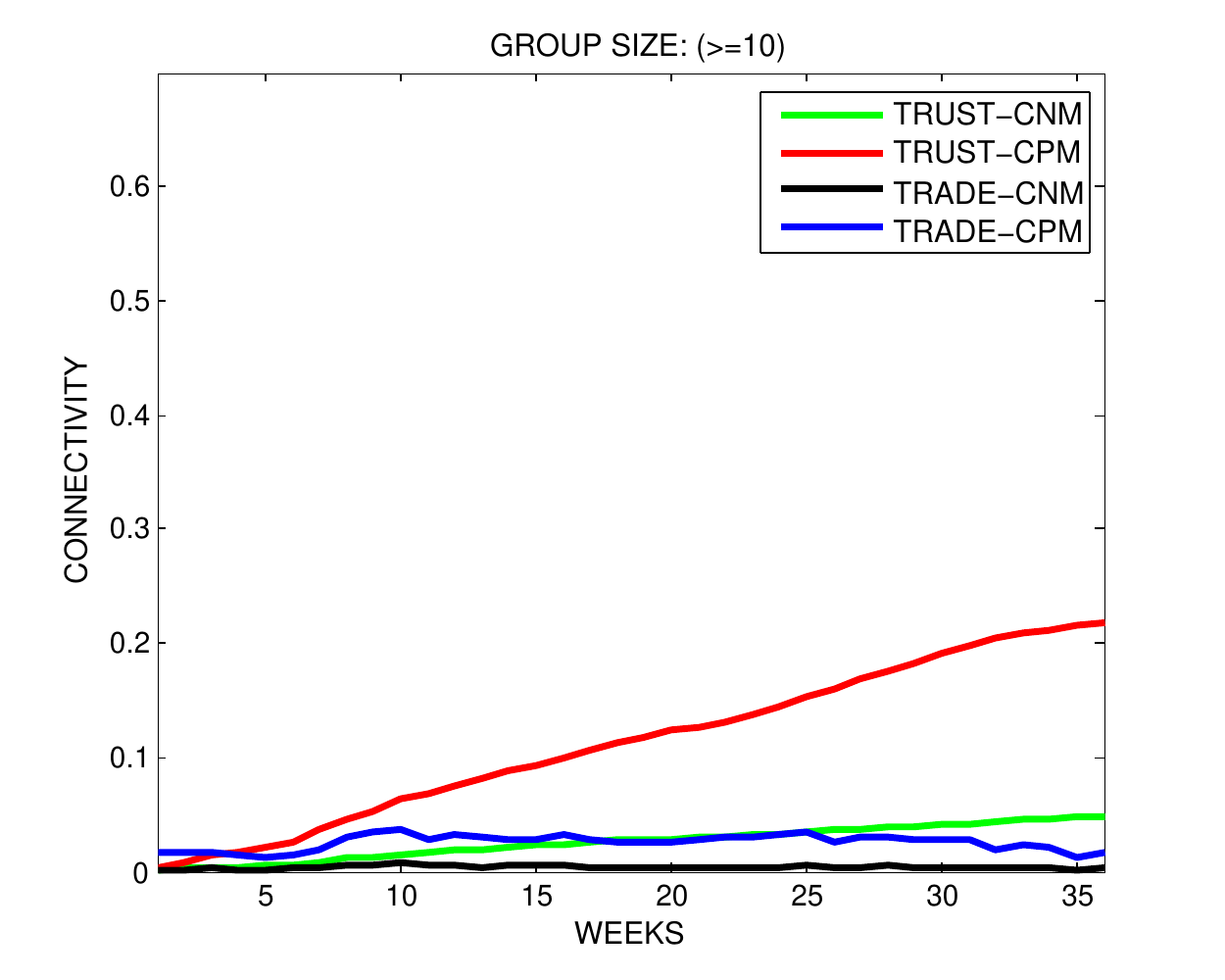}}
				
	\caption{Evolution of trade and trust connectivity within the trust communities(grouped by their sizes)}
	\label{fig:trust_xcorr}
	\squeezeup

	\centering
		\subfloat[]{\includegraphics[width=0.33\textwidth]{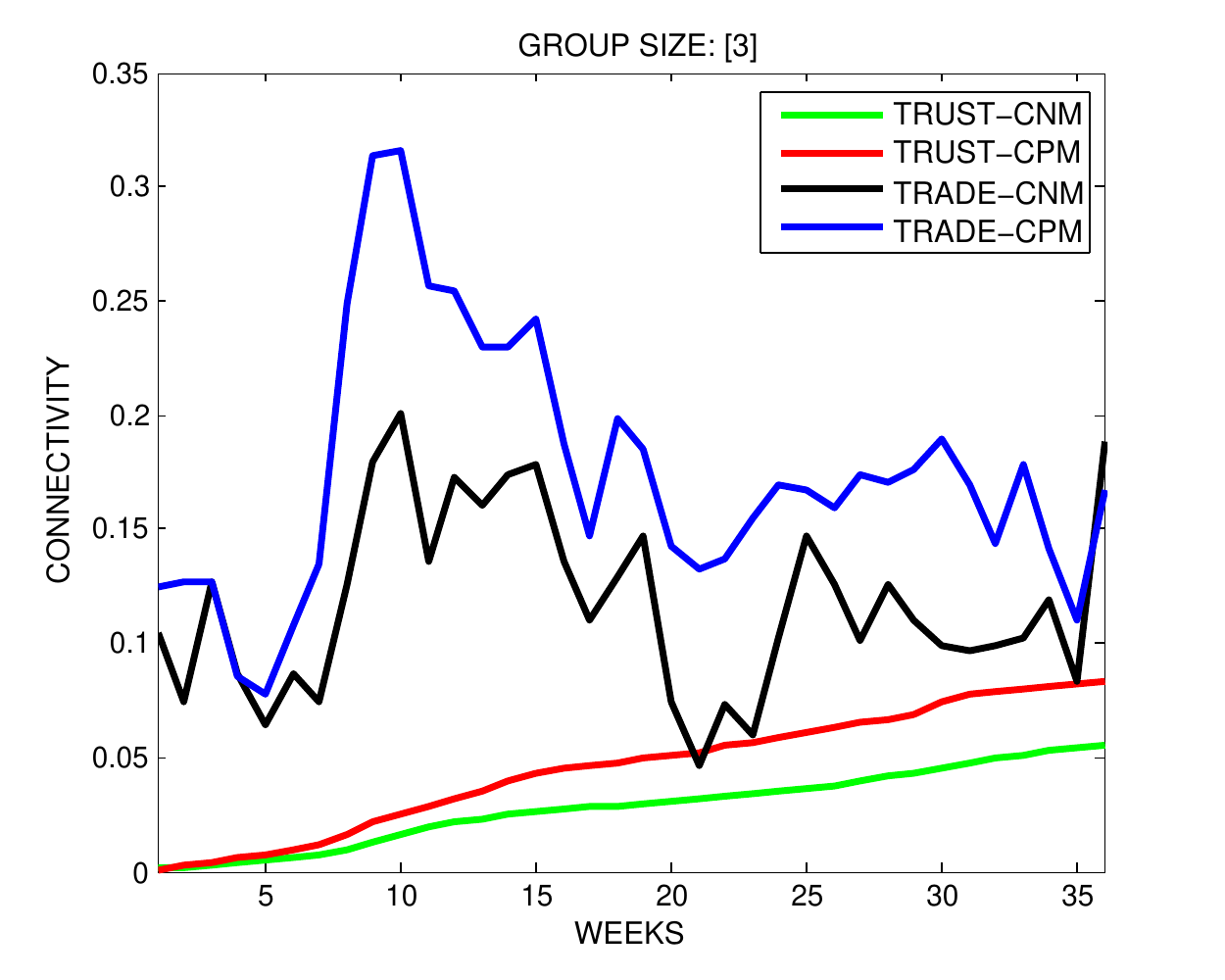}}~\subfloat[]{\includegraphics[width=0.33\textwidth]{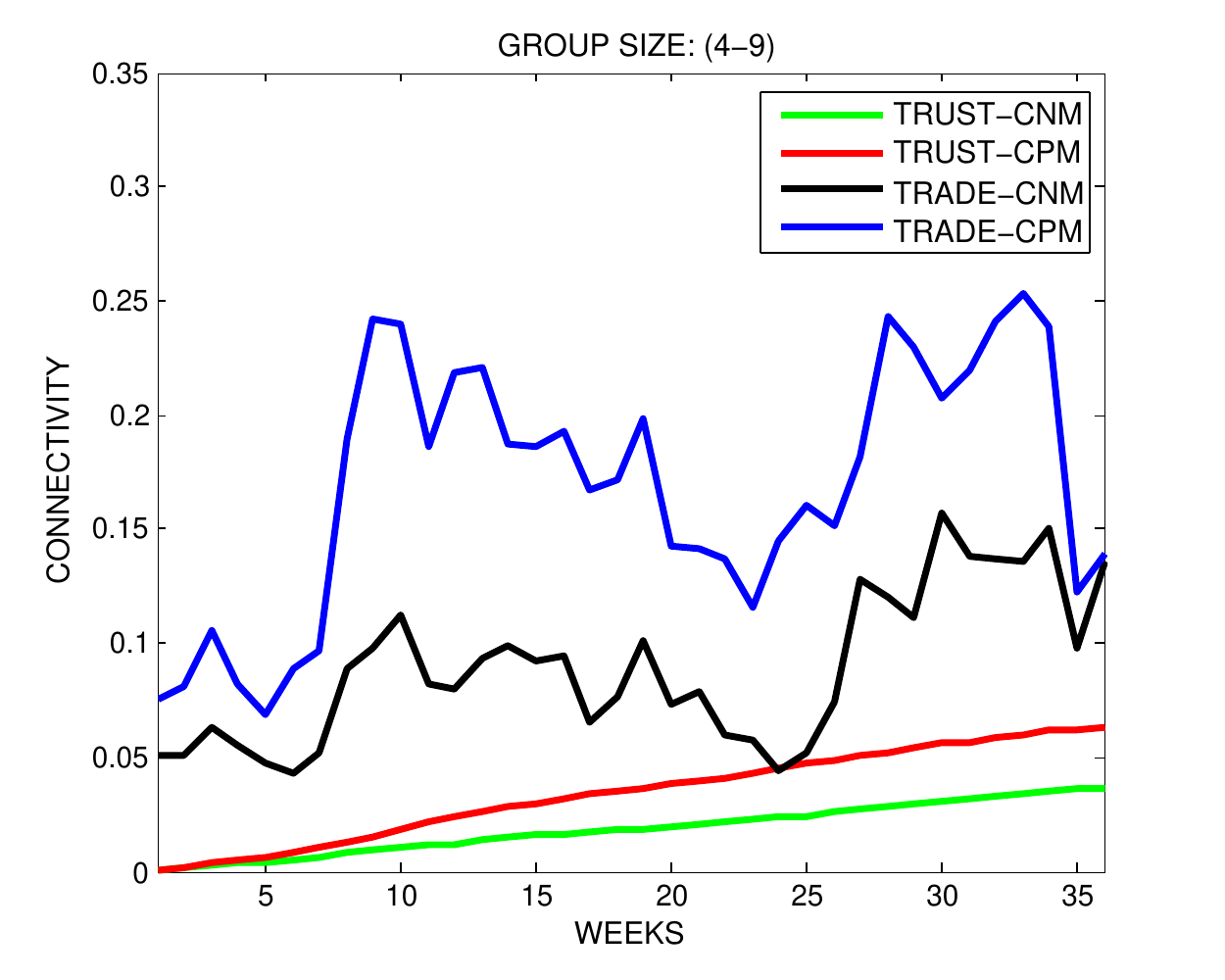}}~\subfloat[]{\includegraphics[width=0.33\textwidth]{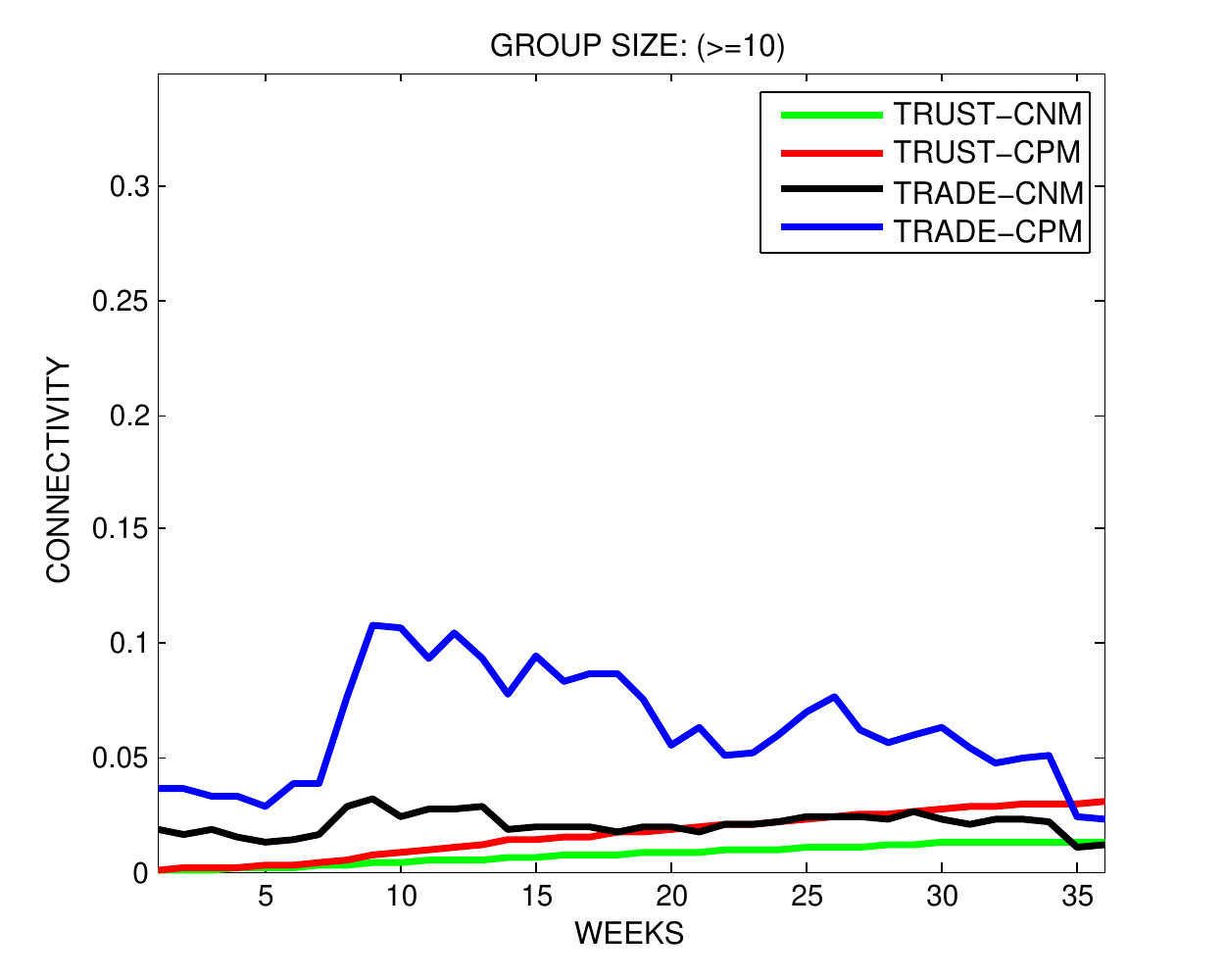}}
					
	\caption{Evolution of trade and trust connectivity within the trade communities (grouped by their sizes)}
	\label{fig:trade_xcorr}
	\squeezeup
\end{figure*}

Figure~\ref{fig:overall_network_connectivity_trust_trade} shows the connectivity of the entire network $G=<V,E>$. As shown in the figure, considering the entire network as single community, shows that the connectivity is of the order $10^{-5}$ throughout the time window. However, this does not represent the actual evolution of connectivity occurring in the network. As mentioned earlier, the entire network is built up from several fragments known as communities. Figures~\ref{fig:trust_xcorr} and \ref{fig:trade_xcorr} show the evolution of average connectivity for the communities build using different algorithms. The average connectivity, at a time stamp $t$, for a specific community detection algorithm is computed as follows:

$\bar{Q}_{\rm trade}^{t,M_i}=\frac{\sum Q_{\rm trade}^{t,k}}{|M_i|}$ $\forall$ $k \in M_i$ and 

$\bar{Q}_{\rm trust}^{t,M_i}=\frac{\sum Q_{\rm trust}^{t,k}}{|M_i|}$ $\forall$ $k \in M_i$

where $M_i$ is the set of communities in the $i^{th}$ category/group and $i \in {1,2,3}$

\subsubsection{Results and discussion}
Figure~\ref{fig:trust_xcorr} shows the average connectivity ($\bar{Q}$) plots for communities based on trust network. Figure~\ref{fig:trust_xcorr}(a) corresponds to $\bar{Q}$ for communities of size $=3$. As shown in the figure, the trust connectivity increases over time whereas the trade connectivity is non-increasing which can be considered as a key take-away. This difference owes to the difference in nature of the trust and trade links; trust is a relationship link whereas trade is an activity link and thus instantaneous. This plot compares the connectivity for CPM defined communities and CNM defined communities for communities of size $3$. From this plot, we find that the maximum trust connectivity reached is approximately $0.56$. This is obtained when communities are defined using CPM definition. The maximum trust connectivity for communities defined by CNM is only $0.33$. Another important observation from this plot is regarding the rate of increase in the trust connectivity. We find that the rate of increase in trust connectivity is $1.7$ times higher for CPM algorithm than for CNM algorithm. We can also see the proportionately higher (approximately $4$ times) trade connectivity for CPM than for CNM communities for the entire time period. 

Figure~\ref{fig:trust_xcorr}(b) shows connectivity plots for communities of size ranging from $4$ to $9$. Here the maximum trust connectivity is approximately $0.43$ for CPM communities whereas the maximum trust connectivity is only $0.23$ for CNM communities. The rate of increase in the trust connectivity is approximately $1.86$ times higher in CPM communities than for CNM communities. Similarly, the trade connectivity is consistently higher ($5$ times on average) in CPM communities over CNM communities. In figure~\ref{fig:trust_xcorr}(c), the plot shows average connectivity for communities of size greater than equal to $10$. As shown in the figure, the maximum trust connectivity in CPM communities is approximately $0.21$ whereas the maximum trust connectivity in CNM communities is only $0.05$. The ratio of rate of increase in trust connectivity in CPM to that in CNM communities is approximately $4$. Unlike smaller communities, the trade connectivity is consistently very low ($0.02$ on average) and negligible for CPM and CNM communities respectively.

Figures~\ref{fig:trade_xcorr}(a),(b),(c) show the average connectivity ($\bar{Q}$) plots for communities based on reduced trade network. In comparison to the trust based communities, we can see that the various connectivity are significantly smaller for all sizes of communities. Figure~\ref{fig:trade_xcorr}(a) shows the plot for trust and trade connectivity for communities of size $3$. As shown in this figure, the trade connectivity is higher on average than the trust connectivity within the communities. The highest trade connectivity is approximately $0.3$  for CPM communities, while the trade connectivity for CNM communities remains lower than that for CPM communities for most of the time period. The ratio of trade connectivity averaged over time between CPM and CNM is $1.63$. For the trust connectivity, the ratio of rate of increase in trust connectivity between CPM and CNM is $1.6$, while the maximum trust connectivity is only $0.08$ for CPM communities. Figure~\ref{fig:trade_xcorr}(b) shows the connectivity for communities of size in range $[4,9]$. As shown in the figure, the general trend of trade connectivity is higher than the trust connectivity. The maximum trust connectivity for CPM communities is $0.2$5 whereas it is only $0.15$ for CNM communities. The trade connectivity in CPM was significantly higher (by approximately $2.5$) for certain windows of time interval. For the trust connectivity, it can be observed that CPM communities have a maximum trust connectivity of $0.06$ whereas the maximum trust connectivity for CNM communities is $0.04$. For communities of size greater than equal to $10$, figure~\ref{fig:trade_xcorr}(c) summarizes the connectivity measures for trust and trade links. As shown in figure, the maximum trade connectivity is as low as $0.1$ for CPM whereas maximum trade connectivity for CNM is only $0.02$. Similarly the maximum trust connectivity for CPM and CNM are $0.035$ and $0.016$ respectively.

To summarize, we find that the various connectivity measures, such as for trade and trust connectivity, are higher when communities are defined using CPM algorithm rather than CNM algorithm. This explains the difference between the different community definitions based on the difference in the community dynamics. Furthermore, we compared the communities derived from the trust network and the reduced trade network. Based on the above mentioned experiments and observation, we find that the trust connectivity is significantly higher in trust based communities than in trade based communities. These results are significant in regards to empirically show the difference between communities which are based on trust relationship and the communities which are based on trade relationship. Although the trade activities are higher in trade based communities, the low trust connectivity corresponds to weaker strength of such communities as will be evident in the following experiments.

\begin{figure*}[t]
	\centering
		\subfloat[]{\includegraphics[width=0.33\textwidth]{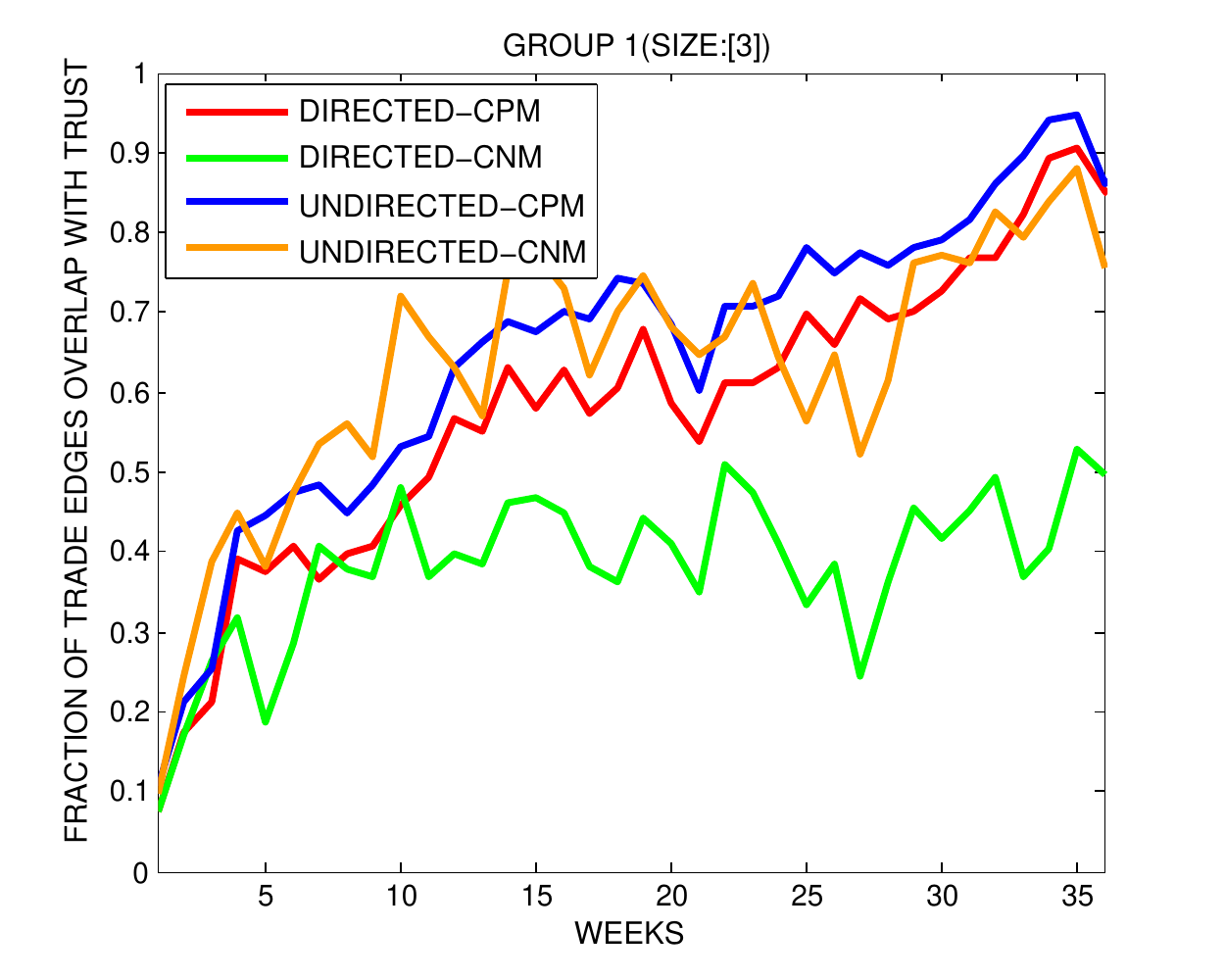}}~\subfloat[]{\includegraphics[width=0.33\textwidth]{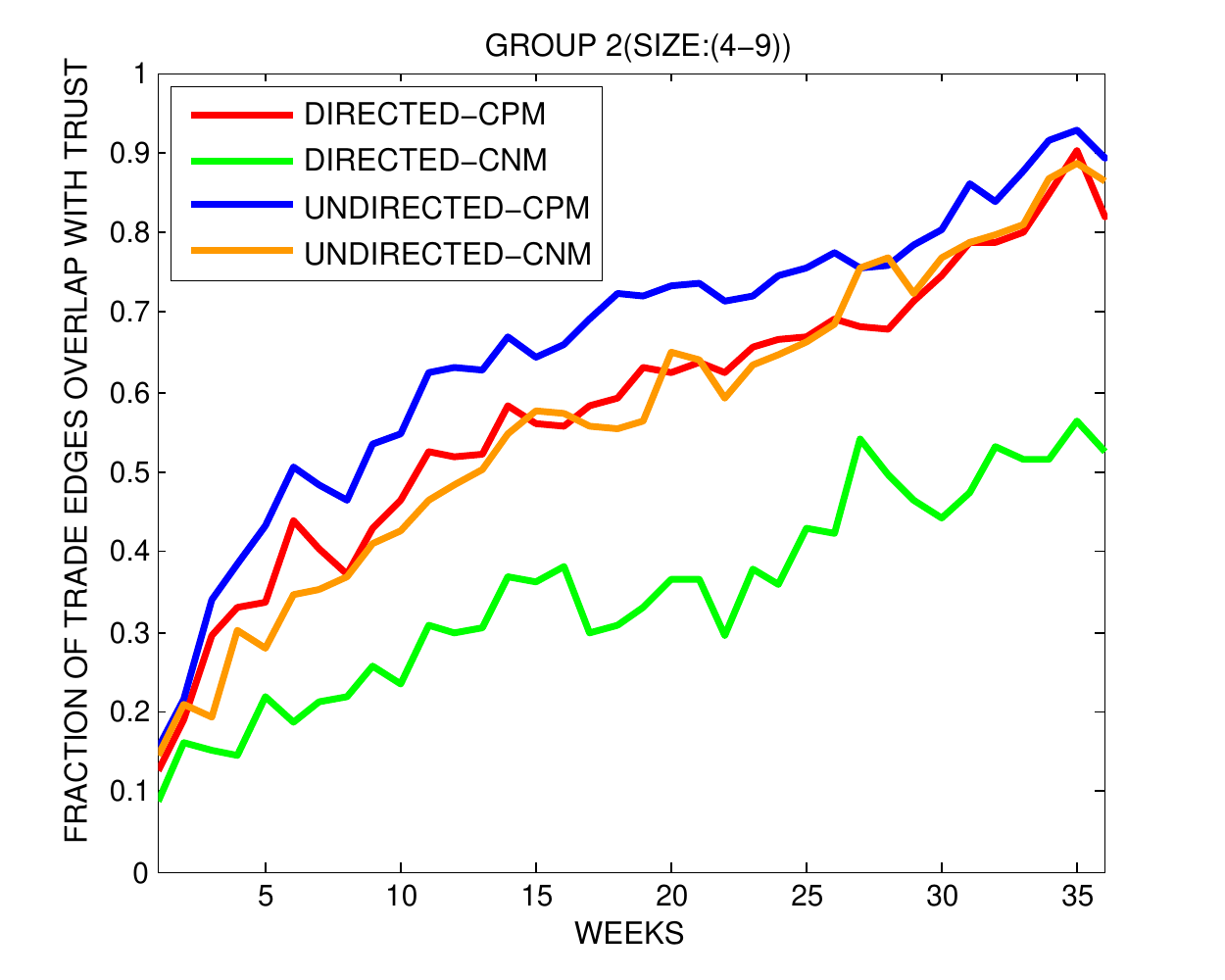}}~\subfloat[]{\includegraphics[width=0.33\textwidth]{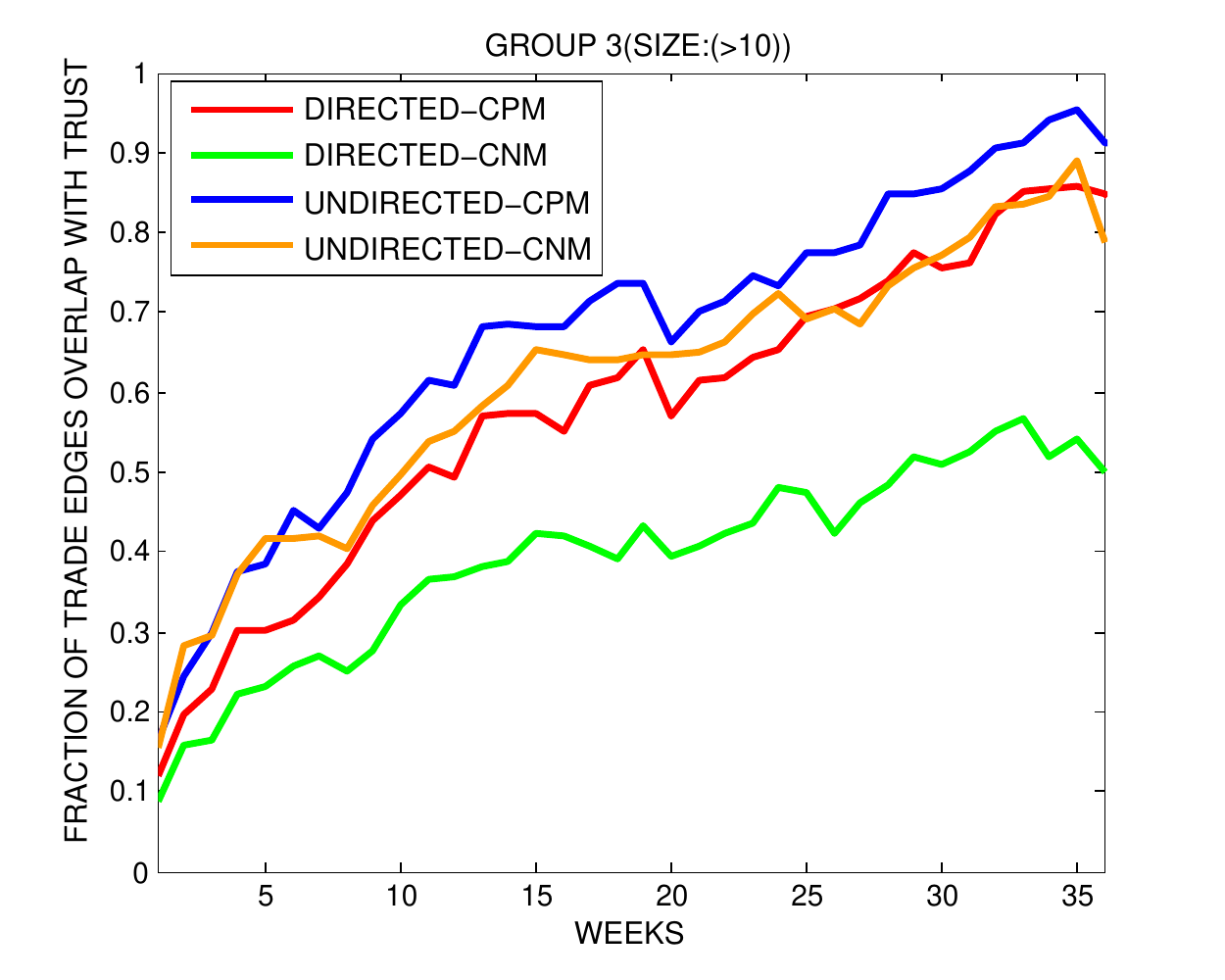}}
			\caption{Analysis of evolution of proportion of trade links overlap with trust links within trust communities.}
	\label{fig:trust_overlap}
	\squeezeup

	\centering
		\subfloat[]{\includegraphics[width=0.33\textwidth]{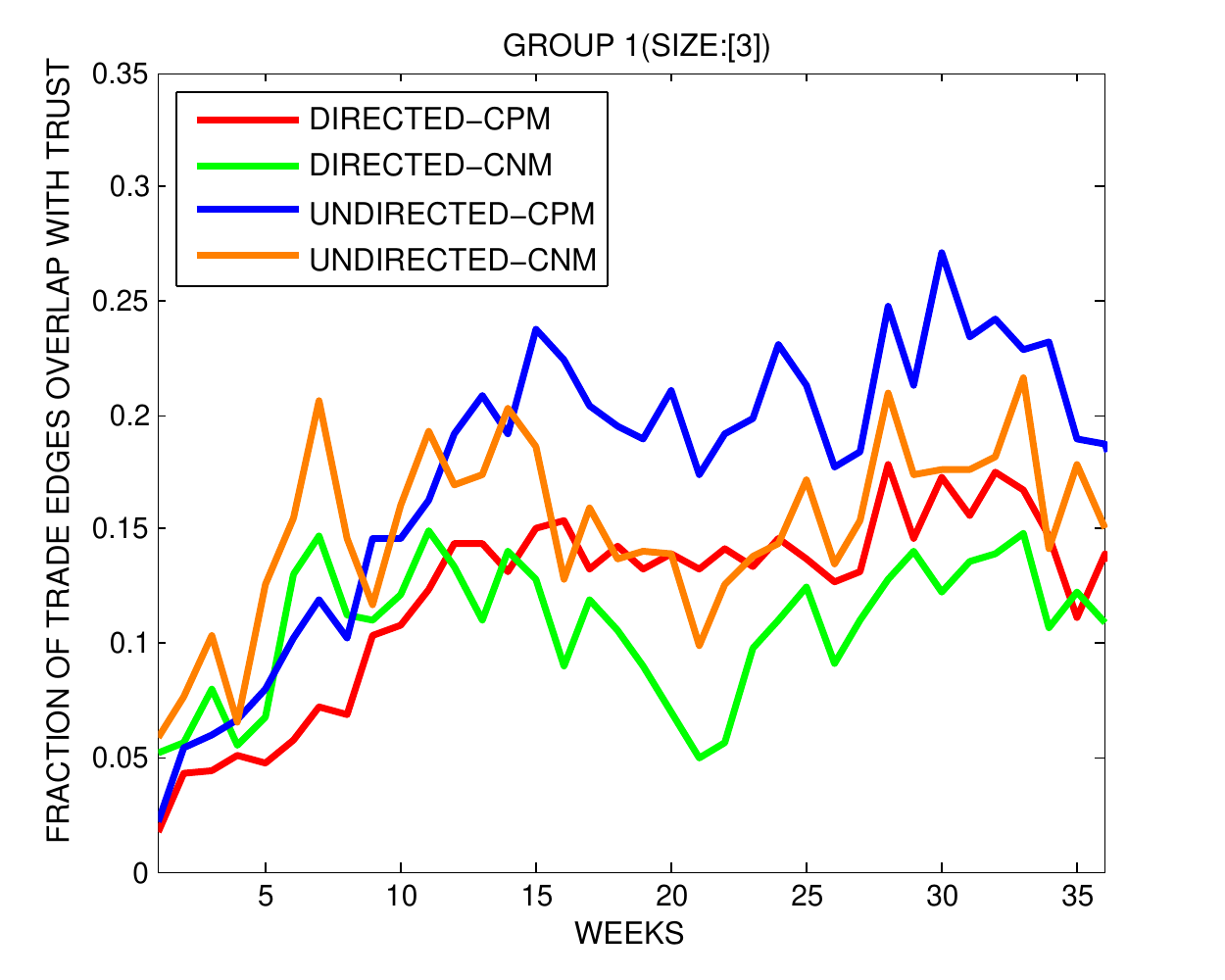}}~\subfloat[]{\includegraphics[width=0.33\textwidth]{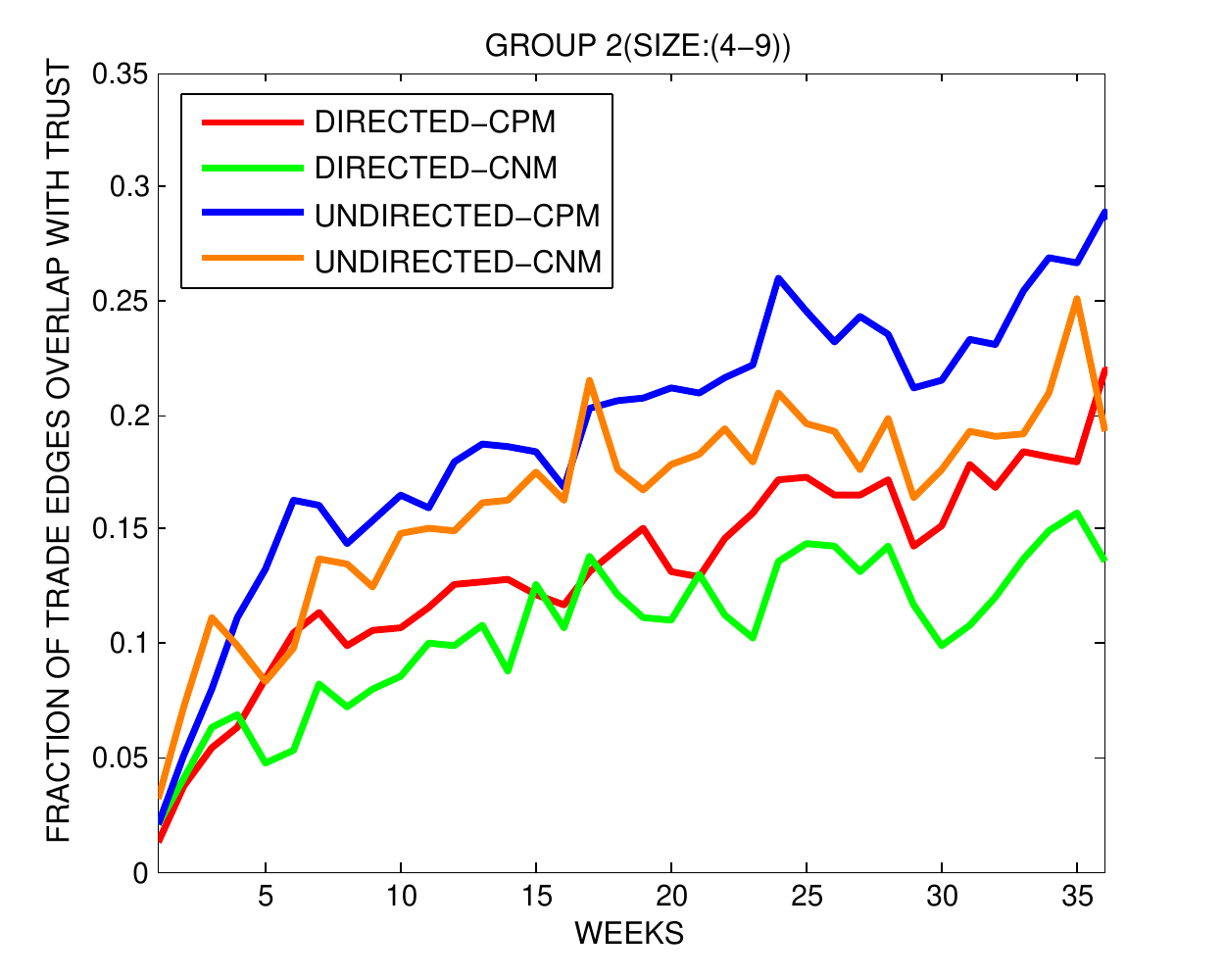}}~\subfloat[]{\includegraphics[width=0.33\textwidth]{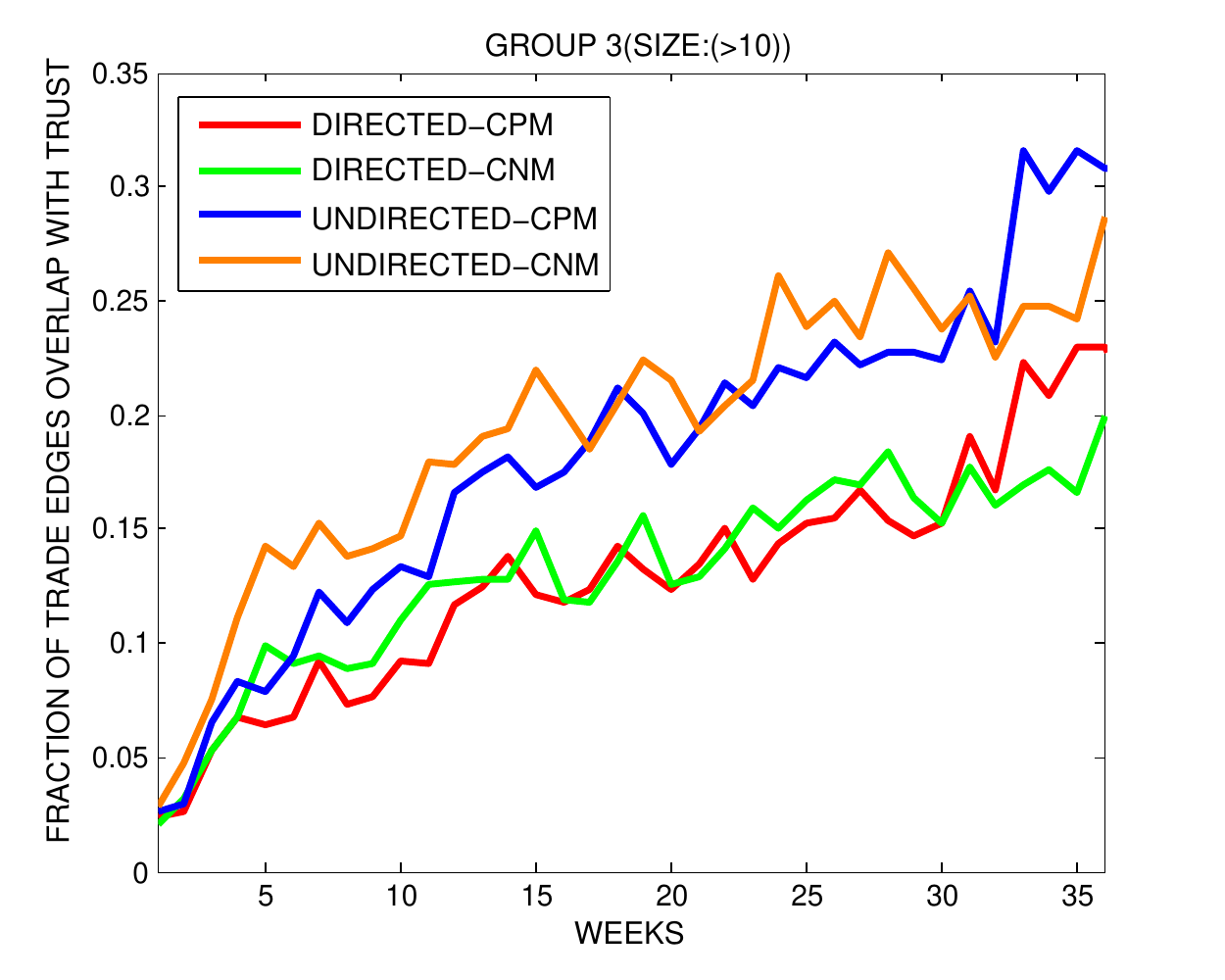}}
\caption{Analysis of evolution of proportion of trade links overlap with trust links within trade communities.}
	\label{fig:trade_overlap}
	\vspace{-1em}
	\squeezeup
\end{figure*}

\subsection{Analysing evolution of activity-relationship overlaps in communities}
In this section, we extend the previous analysis to capture the co-evolutionary aspect of trust and trade within communities. As described in the previous section, the \textit{connectedness} within a community is measured using the connectivity metric. However, this metric captures the trust and the trade connectivity of the communities independently. Capturing interdependence between the trust and trade links within communities is an important and interesting problem. Thus we define metrics to compute overlap between the trust and trade links within the communities. The overlap between trust and trade can be defined in two ways as follows:

\subsubsection{Directed overlap:} Given a snap-shot of a community $k$ at a time-stamp $t$, the directed overlap is the ratio of the directed trade links that overlap with the directed trust links to the total number of directed  trade links in the community $k$ at $t$. It can be mathematically represented as follows:

$O_{\rm directed}^{t,k} = \frac{| \{e \in  E_{\rm trade}^{t,k} | e' = (\rm src(e), \rm dst(e)) \in E_{\rm trust}^t \} |} {| E_{\rm trade}^{t,k} |}$

\subsubsection{Undirected overlap:} Given a snap-shot of a community $k$ at a time-stamp $t$, the undirected overlap is the ratio of the undirected trade links that overlap with the undirected trust links to the total number of trade links in the community $k$ at $t$. It can be mathematically represented as follows:

$O_{\rm undirected}^{t,k} = \frac{| \{ e \in  E_{\rm trade}^{t,k} | e' = \{(\rm src(e), \rm dst(e)) or (\rm dst(e), \rm src(e)) \} \in E_{\rm trust}^t\} |} {| E_{\rm trade}^{t,k} |}$

\subsubsection{Results and discussion}
Figures~\ref{fig:trust_overlap} and \ref{fig:trade_overlap} show the plots for the various overlap metrics of communities. Figures~\ref{fig:trust_overlap}(a),(b),(c) show the overlap metrics for communities derived from the trust network. The \textit{x-axis} in these plots corresponds to time in weeks and the \textit{y-axis} is the overlap metric. Each plot shows $4$ curves: directed CPM corresponding to directed overlap when communities are defined using CPM algorithm, undirected CPM corresponding to undirected overlap for CPM communities, directed CNM corresponding to directed overlap for CNM communities and undirected CNM for undirected overlap in CNM communities. As shown in the figures, the overall trend of all the four metrics is approximately similar for communities of sizes greater than $4$. The overlap trend is, however, slightly different for communities of size $=3$. In figure~\ref{fig:trust_overlap}(a), we find that directed overlap in CNM communities do not have an overall increasing trend. After an initial increase until $7^{th}$ week, the trend is non-increasing unlike directed overlap in CPM communities and undirected overlaps. In comparison to communities of size greater than $3$, we see that the undirected overlap in CNM communities has similar trends as undirected overlap in CPM communities, unlike the trend in communities of size greater than $3$. 

Figures~\ref{fig:trade_overlap}(a),(b),(c) show the overlap metrics for communities derived from reduced trade network. Figure~\ref{fig:trade_overlap}(a) shows the overlap metrics for communities of size $3$. As shown in this figure, all the $4$ overlap metrics assume a non-increasing trend after initial increase until week $7$ or $8$. It can also be seen that the overlap metrics for CNM communities are in general lower than that for CPM communities. In figures~\ref{fig:trade_overlap}(b) and (c) the overlap metrics are shown corresponding to the communities of size in range $[4,9]$ and those with size greater than $9$ respectively. For communities with size in range $[4,9]$ there is an increasing trend for all the metrics and both the directed and undirected overlap for CNM are in general lower than those for CPM communities. The highest overlap is approximately $0.29$ for CPM communities. For larger communities (size greater than $9$), the trend is slightly different. In general, both the directed and undirected overlaps for CNM communities are slightly higher than those for CPM communities though the maximum overlap of $0.31$ is obtained in CPM communities considering undirected overlap between trade and trust.

To summarize, there are two main observations from this experiment. Firstly, we compare the overlap metrics for trust based communities and trade based communities. For the trust based communities the maximum overlap is approximately $95\%$ whereas for the trade based communities the maximum overlap is about $30\%$ only. The significant difference in the overlap of the two is a strong indicator that the communities defined on trade relationship fail to achieve the expected dynamics of a effective community. In general it is expected that strong trust within communities will also derive strong trade activities within the communities as we can see the high percentage of overlap in case of trust based communities. Secondly, we find that very small sized communities (of size $3$) show slightly different overlap dynamics than other communities of larger size.

\subsection{Analysis of communities' strength} 
In this section, we study the evolution of communities when the communities are not treated as isolated units unlike previous experiments. In this multi-relation network, the various communities simultaneously co-exist and engage in various forms of interaction(trust and trade links) with each other. We define metrics to study the evolution of communities when both external(peripheral) and internal(within) interactions are possible for the communities. We have defined a metric to capture this phenomenon. This metric is called the inter to intra link ratio in the communities. Given a snapshot of a community $k$ (within a network $G$) at time stamp $t$, the inter to intra link ratio for the community $k$ is defined in the following manner.

$S_{\rm trade}^{t,k}= \frac{|P_{\rm trade}^{t,k}|}{|E_{\rm trade}^{t,k}|}$

$S_{\rm trust}^{t,k}= \frac{|P_{\rm trust}^{t,k}|}{|E_{\rm trust}^{t,k}|}$

The overall inter-intra link ratio can be computed as follows:

$S_{\rm trade}^{t}=\sum_{k=1}^{K}S_{\rm trade}^{t,k}$

$S_{\rm trust}^{t}=\sum_{k=1}^{K}S_{\rm trust}^{t,k}$

\subsubsection{Results and discussion}
\begin{figure}
	\centering
		\subfloat[Trust based communities.]{\includegraphics[width=0.25\textwidth]{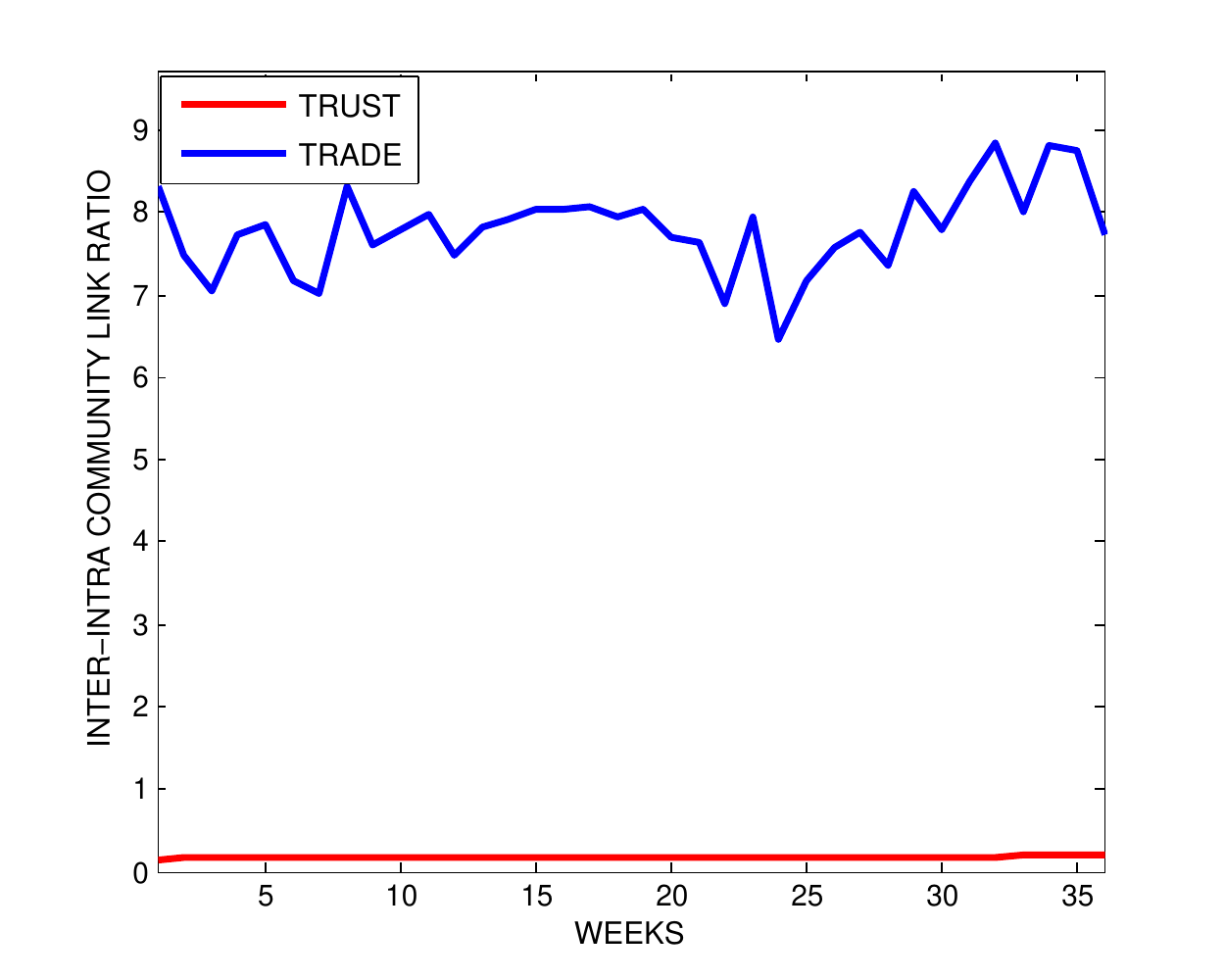}}~\subfloat[Trade based communities.]{\includegraphics[width=0.25\textwidth]{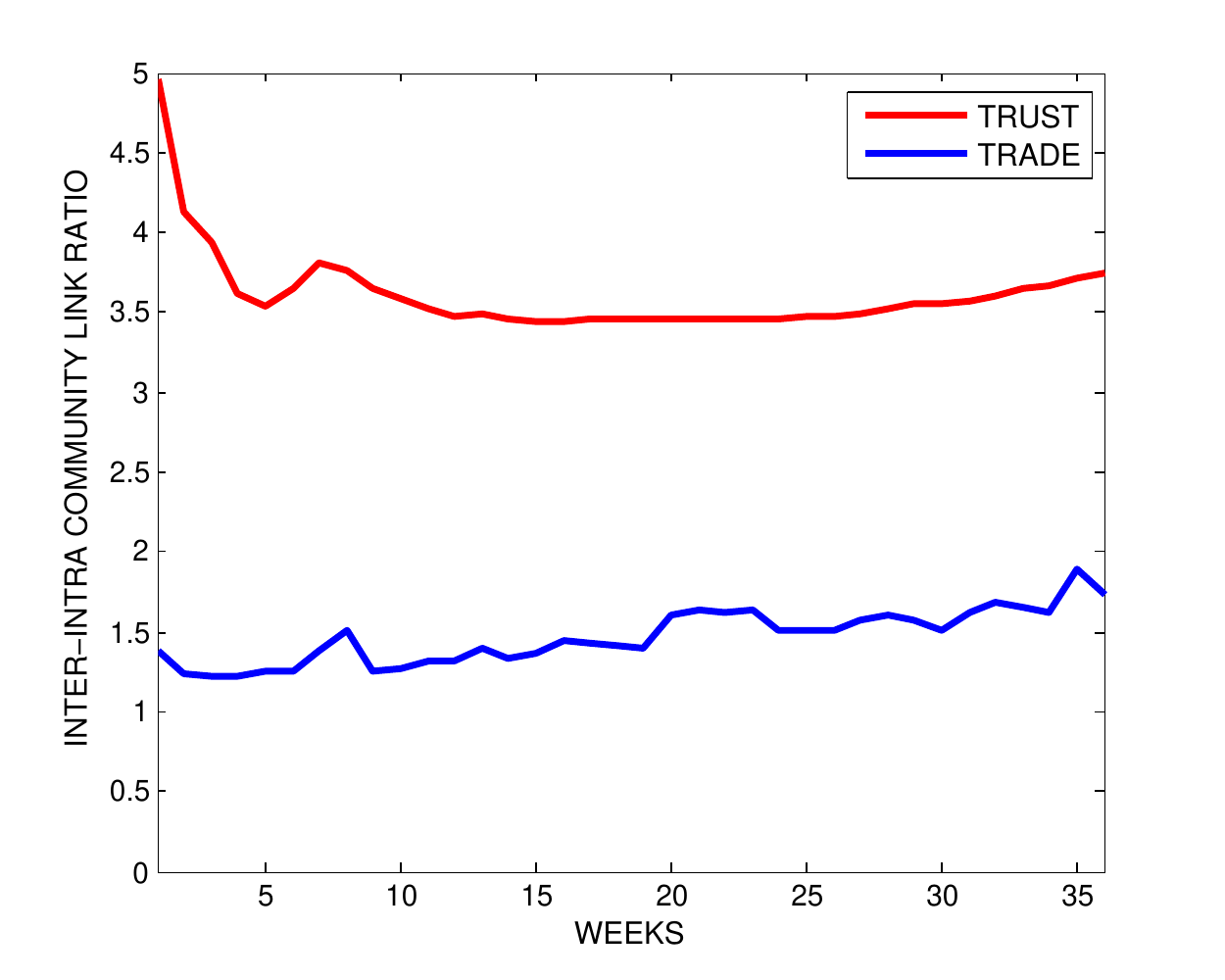}}
			\caption{Analysis of evolution of inter to intra link ratio.}
	\label{fig:trust_trade_inter_intra}
	\squeezeup
\end{figure}

Figures~\ref{fig:trust_trade_inter_intra}(a) and ~\ref{fig:trust_trade_inter_intra}(b) show the evolution of the inter to intra link ratio for communities. Figure~\ref{fig:trust_inter_intra} shows the inter-intra links ratio for trade and trust links for trust based communities. As shown in the figure, the inter-intra trade link ratio is approximately $7.7$ (averaged for the entire time period). High value of inter-intra trade link ratio indicates the inter community trade interactions is $7.7$ times the intra community trade interactions. The ratio fluctuates within $7$ to $9$. However, the inter-intra trust link ratio is as low as$0.1$. Low inter-intra trust link ratio is significant because more trust links are formed within the community than between the communities. It is surprising to find in the trust based communities, the inter community trade is significantly higher in comparison to intra (or within) community trade. 

Figure~\ref{fig:trust_trade_inter_intra} shows the inter-intra links ratio for trade and trust links for trade based communities. As shown in the figure, the inter-intra trade link ratio is approximately $1.25$ in the beginning of the time period and this ratio gradually increases to $1.8$ (approximately) by the end of the time period. Unlike the above case of trust based communities where within trust value was significantly higher, the within trade links for the trade communities are slightly lesser than the inter community trade links. This behavior of excess inter community trade links is attributed to the once in a while trade interaction within communities. Here we see that these once-in-a-while interactions between communities increase over time. The inter-intra trust links ratio is comparatively higher in this case. It is interesting to see that ratio starts as high as $5$ and decline rapidly at the beginning of the time period to nearly $3.5$. Thereafter, this ratio remain persistent with the range of $3.5$ to $3.75$ throughout the time period. It is interesting to see that the inter community trust is higher in the trade based communities. 

To summarize,there are two interesting findings about the inter-intra connectivity of communities. Firstly, for the trust based communities we find that trust links are significantly concentrated within the community whereas the trade links are more significant between the communities than within a community. Secondly, for the trade based communities, we find that the trade links are equally concentrated within and outside a community whereas the trust links are significantly more between the communities than within a trade community. Both these findings signify a opposite role play of trust and trade in terms of inter-intra link dynamics.

%
%

\subsection{Analysing influencers of events in communities}
In this final section of experiments, we discuss the experiment design to estimate the direction of influence in the dynamics of communities. Given the multi-relational nature of the communities and as shown earlier, there are two dynamics happening in any community: trade or trust dynamics. The term `direction of influence' corresponds to a statistical estimate of whether one dynamics is a precursor to another dynamics or if both the dynamics are co-occurring together with statistical significance. 

Although, any magnitude of connectivity can be said as the dynamics of a community, however, for the purpose of this experiment we define dynamics within a community as the occurrence of an event. An event is defined as follows:

\textbf{Event}: An event refers to uncommon dynamics within the community. Given the number of links of a certain $type \in \{\rm trust,\rm trade\}$, $|E_{type}^{t,k}|$, for a community $k$ at $t$, a time series($ts_{\rm type}^k$) of links for $k$ can be constructed by varying $t$ $\in$ $[0,T]$ . An event is said to have occur in the dynamics of the community $k$ if:

$ts_{\rm type}^k(t)-L(ts_{\rm type}^k)>0 \exists t \in [0,T]$

where, $L(ts_{\rm type}^k)=\mu(ts_{\rm type}^k)+\lambda \times \rho(ts_{\rm type}^k)$

$\mu$: mean

$\rho$: standard deviation

$\lambda$: event parameter

The events can be further sub-categorized based on the trust and trade dynamics as follows:

\begin{enumerate}
\item{Sudden jump(trust)}:An event is called sudden jump in trust dynamics when type=`trust' and $\lambda$ $\in$ \{1, 1.5, 2, 2.5\}

\item{Sudden drop(trust)}:An event is called sudden drop in trust dynamics when type=`trust' and $\lambda$ $\in$ \{-1, -1.5, -2, -2.5\}
\item{Sudden jump(trade)}:An event is called sudden jump in trade dynamics when type=`trade' and $\lambda$ $\in$ \{1, 1.5, 2, 2.5\}
\item{Sudden drop(trade)}:An event is called sudden drop in trade dynamics when type=`trade' and $\lambda$ $\in$ \{-1, -1.5, -2, -2.5\}
\end{enumerate}

Tables~\ref{tab:Table_interesting_CNM} and ~\ref{tab:Table_interesting_event_dynamics_CPM} show the distribution of trust communities across the various events. The community definitions being used are CNM and CPM respectively. Similarly, tables~\ref{tab:trade_Table_interesting_CNM} and ~\ref{tab:trade_Table_interesting_event_dynamics_CPM} show the distribution of trade communities across the four events. The community definitions being used are CNM and CPM respectively. These tables give an estimate of the probability of the occurrence of an event. 

Given the information about the events in communities, in the next step we select communities for two types of analysis: precursor direction (direction of influence) for sudden jump events and precursor direction(direction of influence) for sudden drop events. In order to determine direction of influence in either of the above two cases, we select only those communities in which both trade and trust events occur anywhere in the time period $[0,T]$. 

\begin{table}
	\centering
	\caption{Table showing the statistics about the different types of interesting events occurring in the \textbf{CNM} based \textbf{trust} communities.}
	\scalebox{0.7}{
		\begin{tabular}{|l||l|l|l|l|}
		\hline
		\textbf{Events }& \textbf{$\lambda$=1.0 }& \textbf{$\lambda$=1.5} & \textbf{$\lambda$=2.0} & \textbf{$\lambda$=2.5}\\
		\hline
		\hline
		\textbf{Sudden drop} & 179(6.20\%) & 116(4.02\%) & 70(2.42\%) & 43(1.49\%)\\
		\textbf{(trust)}&&&&\\
		\hline
		\textbf{Sudden jump}& 2883(99.79\%) & 2883(99.79\%) & 2882(99.76\%) & 2854(98.79\%)\\
		\textbf{(trust)}&&&&\\
		
		\hline
		\textbf{Sudden drop} & 128(4.43\%) & 33(1.14\%) & 5(0.17\%) & 0(0\%)\\
		\textbf{(trade)}&&&&\\
		\hline
		\textbf{Sudden jump} & 1555(53.82\%) & 1555(53.82\%) & 1542(53.37\%) & 1487(51.47\%)\\
		\textbf{(trade)}&&&&\\
		\hline
			
		\end{tabular}
	}
	\vspace{-1em}
	\label{tab:Table_interesting_CNM}
	\squeezeup
\end{table}

\begin{table}
	\centering
	\caption{Table showing the statistics about the direction of influence in the \textbf{CNM} based \textbf{trust} communities which experienced interesting event (\textbf{sudden jump}) in both trade and trust dynamics.}
	\scalebox{0.75}{
		\begin{tabular}{|l||l|l|l|l|}
		\hline
		\textbf{Precursor}& \textbf{$\lambda$=1.0 }& \textbf{$\lambda$=1.5} & \textbf{$\lambda$=2.0} & \textbf{$\lambda$=2.5}\\
		\hline
		\hline
		\textbf{Trade} & 551 (35.50\%) & 542(34.92\%) & 537(34.92\%) & 537(36.63\%)\\
		\hline
		\textbf{Trust}& 494(31.83\%) & 532(34.28\%) & 542(35.24\%) & 507(34.58\%)\\
		
		\hline
		\textbf{Co-occurrence} & 507(32.67\%) & 478(30.80\%) & 459(29.84\%) &422(28.79\%)\\
		\hline
		\textbf{Total } & 1552 & 1552 & 1538 & 1466\\
		\textbf{communities}&&&&\\
		\hline
			
		\end{tabular}
	}
	\vspace{-1em}
	\label{tab:Table_influencer_CNM_sudden_jump2}
	\squeezeup
\end{table}

\subsubsection{Results and discussion}
We discuss these results in two parts. First we discuss the findings for the trust based communities and then we discuss the findings of the trade based communities. As described earlier, the trust based communities can be defined  using (1)CNM or (2)CPM algorithms. We first discuss the results for CNM based trust communities. Table~\ref{tab:Table_interesting_CNM} gives statistics for the above mentioned unusual events within the communities. From this table, we find that sudden jump events in trust dynamics happen in nearly all the communities $(99-98\%)$. The sudden jump event in trade dynamics is the next significant event occurring in approximately $(51-53\%)$ of communities. Sudden drop event in trust dynamics is a rare event$(1.5-6\%)$ and fewer communities $(1.5\%)$ are found with this event as the threshold($\lambda$) is increased from $1$ to $2.5$. Moreover the sudden drop event occurs very rarely $(> 4\%)$ and there is no community with this event when $\lambda = 2.5$. In table~\ref{tab:Table_influencer_CNM_sudden_jump2}, we describe the relationship between the various unusual events within the communities. As shown in this table, the total community rows describe the count of communities when both the sudden jump in trade dynamics and sudden jump in trust dynamics happen in a community during its evolution. The precursor/ influencer relationship describe the order of occurrence of the trade and trust unusual events within a community. The first row describes the count of the communities in which the trade event occurred before the trust event during the its evolution. We find that approximately $35-36.6\%$ of communities first experienced a unusual trade dynamics followed by a unusual trust dynamics. While the trust event was precursor to trade event for about $32-35\%$ of the communities, we also find that $29-32.6\%$ of communities experienced both an unusual trade and trust dynamics simultaneously. From this table we find that trade event is a precursor of trust event along the same lines as was reported in \cite{roy13} with a slightly higher probability than the opposite case though the difference is not significant in this case. Table~\ref{tab:Table_influencer_CNM_sudden_drop2} shows the relationship between the unusual drop in trust and trade dynamics of a community. Similar to the sudden jump event (table~\ref{tab:Table_influencer_CNM_sudden_jump2}), we find the precursor of unusual drop event in communities. As shown in the table, a total of $80$ communities experienced both a sudden drop in trade and a sudden drop in trust for $\lambda = 1$ and only $14$ communities have both the events when $\lambda = 1.5$. For higher values of $\lambda$ we do not find any communities with such characteristics. Here in this table we find that for lower values of $\lambda (= 1.0)$ trade is the precursor while for $\lambda = 1.5$ trust is precursor. Thus it is hard to find any consistency about any precursor event in case of unusual drop event in CNM based trust communities.

\begin{table}
	\centering
	\caption{Table showing the statistics about the direction of influence in the \textbf{CNM} based \textbf{trust} communities which experienced interesting event (\textbf{sudden drop}) in both trade and trust dynamics.}
	\scalebox{0.75}{
		\begin{tabular}{|l||l|l|}
		\hline
		\textbf{Precursor}& \textbf{$\lambda$=1.0 }& \textbf{$\lambda$=1.5} \\
		\hline
		\hline
		\textbf{Trade} & 49(61.25\%) & 4(28.6\%) \\
		\hline
		\textbf{Trust}& 24(30\%) & 9(64.4\%) \\
		
		\hline
		\textbf{Co-occurrence} & 7(8.75\%) & 1(7\%)\\
		\hline
		\textbf{Total } & 80 & 14 \\
		\textbf{communities}&&\\
		\hline
			
		\end{tabular}
	}
	\vspace{-1em}
	\label{tab:Table_influencer_CNM_sudden_drop2}
	\squeezeup
\end{table}

Table~\ref{tab:Table_interesting_event_dynamics_CPM} summarizes the statistics for the unusual events in the CPM defined trust communities. As shown in this table, sudden jump is trade dynamics is the most significant event across all $\lambda$ values. Almost $69-72.6\%$ of total communities experience this event. The sudden drop in trust dynamics and sudden jump in trust dynamics are significant for $\lambda = 1.0$ and fewer communities experience these events as $\lambda$ is increased. These fractions range from $6\%$ to $10\%$. We also find that sudden drop in trade dynamics is again a rare event because less than $3.6\%$ of total communities experience this event. In table~\ref{tab:Table_influencer_CPM_sudden_jump}, we describe relationship between the trade and trust sudden jump events occurring in the communities. The third row of this table shows the count of communities for which both sudden jump in trade and sudden jump in trust occurs during their evolution. Unlike CNM defined communities, we find a stronger indicator of precursor from this table. Across all the values of $\lambda$, we find that in $73.5-83.5\%$ of communities with both trade and trust event, the sudden jump in trade dynamics is followed by a sudden jump in trust dynamics. From this table we can clearly see that unusual trait in trust behaviour in a community is preceded by a unusual trait in trade behaviour within a community. Although we find a similar behavior in case of CNM defined communities, this behavior is strongly manifested in case of CPM defined communities which by definition are more tightly knit than CNM defined communities. 
\begin{table}
	\centering
	\caption{Table showing the statistics about the different types of interesting events occurring in the \textbf{CPM} based \textbf{trust} communities. }
	\scalebox{0.75}{
		\begin{tabular}{|l||l|l|l|l|}
		\hline
		\textbf{Events }& \textbf{$\lambda$=1.0 }& \textbf{$\lambda$=1.5} & \textbf{$\lambda$=2.0} & \textbf{$\lambda$=2.5}\\
		\hline
		\hline
		\textbf{Sudden drop} & 1887(65.6\%) & 1136(39.5\%) & 539(18.7\%) & 287(9.9\%)\\
		\textbf{(trust)}&&&&\\
		\hline
		\textbf{Sudden jump}& 1785(62\%) & 1027(35.7\%) & 555(19.3\%) & 282(9.8\%)\\
		\textbf{(trust)}&&&&\\
		
		\hline
		\textbf{Sudden drop} & 104(3.6\%) & 9(0.3\%) & 0(0.0\%) & 0(0.0\%)\\
		\textbf{(trade)}&&&&\\
		\hline
		\textbf{Sudden jump} & 2089(72.6\%) & 2089(72.6\%) & 2075(72\%) & 1992(69\%)\\
		\textbf{(trade)}&&&&\\
		\hline
			
		\end{tabular}
	}
	\vspace{-1em}
	\label{tab:Table_interesting_event_dynamics_CPM}
	\squeezeup
\end{table}

\begin{table}
	\centering
	\caption{Table showing the statistics about the direction of influence in the \textbf{CPM} based \textbf{trust} communities which experienced interesting event (\textbf{sudden jump}) in both trade and trust dynamics.}
	\scalebox{0.75}{
		\begin{tabular}{|l||l|l|l|l|}
		\hline
		\textbf{Precursor}& \textbf{$\lambda$=1.0 }& \textbf{$\lambda$=1.5} & \textbf{$\lambda$=2.0} & \textbf{$\lambda$=2.5}\\
		\hline
		\hline
		\textbf{Trade} & 1127(83.5\%) & 647(83.8\%) & 315(79.5\%) & 136(73.5\%)\\
		\hline
		\textbf{Trust}& 72(5.5\%) & 29(3.8\%) & 16(4\%) & 12(6.5\%)\\
		
		\hline
		\textbf{Co-occurrence} & 150(11\%) & 96(12.4\%) & 65(16.5\%) & 37(20\%)\\
		\hline
		\textbf{Total } & 1349 & 772 & 396 & 185\\
		\textbf{communities}&&&&\\
		\hline
			
		\end{tabular}
	}
	\vspace{-1em}
	\label{tab:Table_influencer_CPM_sudden_jump}
	\squeezeup
\end{table}


		%
			%


\begin{table}
	\centering
	\caption{Table showing the statistics about the different types of interesting events occurring in the \textbf{CNM} based \textbf{trade} communities.}
	\scalebox{0.75}{
		\begin{tabular}{|l||l|l|l|l|}
		\hline
		\textbf{Events }& \textbf{$\lambda$=1.0 }& \textbf{$\lambda$=1.5} & \textbf{$\lambda$=2.0} & \textbf{$\lambda$=2.5}\\
		\hline
		\hline
		\textbf{Sudden drop} & 90(4.34\%) & 63(3.04\%) & 42(2.03\%) & 30(1.45\%)\\
		\textbf{(trust)}&&&&\\
		\hline
		\textbf{Sudden jump}& 829(39.99\%) & 829(39.99\%) & 825(39.80\%) & 810(39.07\%)\\
		\textbf{(trust)}&&&&\\
		
		\hline
		\textbf{Sudden drop} & 108(5.21\%) & 35(1.69\%) & 2(0.096\%) & 0(0\%)\\
		\textbf{(trade)}&&&&\\
		\hline
		\textbf{Sudden jump} & 2073(100\%) & 2073(100\%) & 2063(99.52\%) & 1993(96.14\%)\\
		\textbf{(trade)}&&&&\\
		\hline
			
		\end{tabular}
	}
	\vspace{-1em}
	\label{tab:trade_Table_interesting_CNM}
	\squeezeup
\end{table}

We now discuss the results for communities derived from trade network. Following the same pattern as for the results for trust based communities; we first discuss the results for CNM defined communities. Table~\ref{tab:trade_Table_interesting_CNM} shows the statistics for various interesting events occurring in CNM defined communities. As shown in this table, sudden jump in trade happens in nearly all of the communities $(96-100\%)$. Sudden jump in trust occurs in approximately $40\%$ of the communities across all values of $\lambda$. Sudden drop in trust and trade are rare events in communities. Table~\ref{tab:Table_influencer_CNM_sudden_jump} describes the relationship between sudden jump in trade and sudden jump in trust for the communities which experiences both the events in course of their evolution. As shown in the table, for $\lambda=1.0,1.5$ trade event seems to be precursor for trust event in communities. But as the $\lambda$ is increased to $2.0$ and $2.5$ the relationship reverse and trust events shows precedence over trade events. From this table it is hard to draw conclusion about the direction of influence between unusual trust dynamics and unusual trade dynamics within communities. Table~\ref{tab:Table_influencer_CNM_sudden_drop} describes the relationship between sudden drop trade event and sudden drop trust event in communities. Similar to the sudden jump case, the sudden drop events in CNM defined communities has $\lambda$ dependent relationship between sudden drop trade and sudden drop trust within communities.


\begin{table}
	\centering
	\caption{Table showing the statistics about the direction of influence in the \textbf{CNM} based \textbf{trade} communities which experienced interesting event (\textbf{sudden jump}) in both trade and trust dynamics.}
	\scalebox{0.75}{
		\begin{tabular}{|l||l|l|l|l|}
		\hline
		\textbf{Precursor}& \textbf{$\lambda$=1.0 }& \textbf{$\lambda$=1.5} & \textbf{$\lambda$=2.0} & \textbf{$\lambda$=2.5}\\
		\hline
		\hline
		\textbf{Trade} & 367 (44.27\%) & 322(35\%) & 303(36.55\%) & 259(31.74\%)\\
		\hline
		\textbf{Trust}& 217(26.18\%) & 261(29\%) & 286(34.50\%) & 286(35.0\%)\\
		
		\hline
		\textbf{Co-occurrence} & 245(29.55\%) & 246(29.67\%) & 227(27.38\%) & 204(25.00\%)\\
		\hline
		\textbf{Total } & 829 & 829 & 816 & 749\\
		\textbf{communities}&&&&\\
		\hline
			
		\end{tabular}
	}
	\label{tab:Table_influencer_CNM_sudden_jump}
	\squeezeup
\vspace{1em}
	\centering
	\caption{Table showing the statistics about the direction of influence in the \textbf{CNM} based \textbf{trade} communities which experienced interesting event (\textbf{sudden drop}) in both trade and trust dynamics.}
	\scalebox{0.75}{
		\begin{tabular}{|l||l|l|l|}
		\hline
		\textbf{Precursor}& \textbf{$\lambda$=1.0 }& \textbf{$\lambda$=1.5} & \textbf{$\lambda$=2.0} \\
		\hline
		\hline
		\textbf{Trade} & 32(71.1\%) & 6(46.15\%) & 0(0.0\%)\\
		\hline
		\textbf{Trust}& 9(20\%) & 5(38.46\%) & 1 (100\%) \\
		
		\hline
		\textbf{Co-occurrence} & 4(8.9\%) & 2(15.38\%) & 0(0.0\%)\\
		\hline
		\textbf{Total } & 45 & 13 & 1 \\
		\textbf{communities}&&&\\
		\hline
			
		\end{tabular}
	}
	\vspace{-1em}
	\label{tab:Table_influencer_CNM_sudden_drop}
	\squeezeup
\vspace{1em}
	\centering
	\caption{Table showing the statistics about the different types of interesting events occurring in the \textbf{CPM} based \textbf{trade} communities. These interesting events are based on the dynamics of trade and trust within these communities.}
	\scalebox{0.7}{
		\begin{tabular}{|l||l|l|l|l|}
		\hline
		\textbf{Events }& \textbf{$\lambda$=1.0 }& \textbf{$\lambda$=1.5} & \textbf{$\lambda$=2.0} & \textbf{$\lambda$=2.5}\\
		\hline
		\hline
		\textbf{Sudden drop} & 192(6.6\%) & 172(5.9\%) & 135(4.65\%) & 107(3.7\%)\\
		\textbf{(trust)}&&&&\\
		\hline
		\textbf{Sudden jump}& 1274(44\%) & 1274(44\%) & 1269(43.8\%) & 1245(43\%)\\
		\textbf{(trust)}&&&&\\
		\hline
		\textbf{Sudden drop(trade)} & 149(5\%) & 23(0.8\%) & 2(0.07\%) & 1(0.0\%)\\
		\textbf{(trade)}&&&&\\
		\hline
		\textbf{Sudden jump(trade)} & 2898(100\%) & 2897(99.9\%) & 2879(99\%) & 2756(95\%)\\
		\textbf{(trade)}&&&&\\
		\hline
			
		\end{tabular}
	}
	\vspace{-1em}
	\label{tab:trade_Table_interesting_event_dynamics_CPM}
	\squeezeup
	\end{table}
Table~\ref{tab:trade_Table_interesting_event_dynamics_CPM} summarizes the statistics about the unusual events in CPM defined communities. As shown in the table, the sudden jump event in trade dynamics occurs in nearly $95-100\%$ of communities depending on the values of $\lambda$. Sudden jump in trust occurs in nearly $43-44\%$ of communities. The sudden drops events in trust and trade dynamics are very rare events. Table~\ref{tab:Table_influencer_CPM_sudden_jump} describes the relationship between sudden jump in trust dynamics and sudden jump in trade dynamics of communities. In this table we can see that the sudden jump in trade dynamics precedes the sudden jump in trust dynamics across all the $\lambda$ values. This table shows that for communities which experiences sudden jump in trade and sudden jump in trust, the sudden jump in trade is followed by a sudden jump in trust. We find that the precursor relationship is distinctly manifested in case of tightly knit CPM communities in comparison to CNM communities. 

	\begin{table}
\vspace{1em}
	\centering
	\caption{Table showing the statistics about the direction of influence in the \textbf{CPM} based \textbf{trade} communities which experienced interesting event (\textbf{sudden jump}) in both trade and trust dynamics.}
	\scalebox{0.7}{
		\begin{tabular}{|l||l|l|l|l|}
		\hline
		\textbf{Precursor }& \textbf{$\lambda$=1.0 }& \textbf{$\lambda$=1.5} & \textbf{$\lambda$=2.0} & \textbf{$\lambda$=2.5}\\
		\hline
		\hline
		\textbf{Trade} & 690(54\%) & 616(48.4\%) & 562(44.9\%) & 490(42.8\%)\\
		\hline
		\textbf{Trust}& 318(25\%) & 392(30.8\%) & 442(35.3\%) & 416(36.4\%)\\
		
		\hline
		\textbf{Co-occurrence} & 266(21\%) & 265(20.8\%) & 249(19.8\%) & 238(20.8\%)\\
		\hline
		\textbf{Total } & 1274 & 1273 & 1253 & 1144\\
		\textbf{communities}&&&&\\
		\hline
			
		\end{tabular}
	}
	\label{tab:Table_influencer_CPM_sudden_jump}
	\squeezeup

		%
			%
\end{table}

\section{Conclusions}
In this paper we have studied an interesting problem of evolution in multi-relational social networks. Unlike conventional approaches to study evolution in networks, we analyze the evolution by fragmenting the larger network into macro units known as communities. In a multi-relational setting, there are several interesting phenomena about the trade and trust dynamics which were studied in this work. Specifically, we find that the trust based communities are strongly connected than the trade based ones. Another essential difference between the dynamics of trust based communities and trade based communities is the higher co-evolution rate in trust based communities. Averaging for the overall network, we find that in general  a community has only one type of links strongly concentrated within itself while the other type of links are distributed outside the community. For tightly knit communities, we find that in approximately $70\%$ of communities experience an unusual behavior in trade dynamics (sudden jump in trade activity) also experience an unusual behavior in their trust dynamics in the later period.





%

\bibliographystyle{abbrv}
\bibliography{ref}

\end{document}